\RequirePackage{fix-cm}
\documentclass[twocolumn,epjc3]{svjour3}  
\smartqed  
\RequirePackage{graphicx}
\usepackage{epsfig}
\usepackage{amsmath}
\usepackage{amssymb}
\usepackage{amsbsy}
\usepackage{enumerate}
\usepackage{url}
\usepackage{graphicx}
\usepackage{xspace}
\usepackage{cite}
\usepackage{fancyvrb}
\usepackage{booktabs}
\newcommand\new{\newcommand}         

\def\beq{\begin{equation}}   
\def\eeq{\end{equation}}
\def\bea{\begin{eqnarray}}  
\def\eea{\end{eqnarray}} 
\newcommand{\bite}{\begin{itemize}}
\newcommand{\eite}{\end{itemize}}
\newcommand{\bcen}{\begin{center}}
\newcommand{\ecen}{\end{center}}

\def\Ac{A_{\mathrm{t\bar{t}}}^{\mathrm{C}}}

\new{\eV}   {{\ifmmode {\mathrm{ eV}}\else ${\mathrm{ eV}}$\fi}}
\new{\MeV}  {{\ifmmode {\mathrm{ MeV}}\else ${\mathrm{ MeV}}$\fi}}
\new{\MeVc} {{\ifmmode {\mathrm{ MeV}}/c\else ${\mathrm{ MeV}}/c$\fi}}
\new{\MeVcc}{{\ifmmode {\mathrm{ MeV}}/c^2\else ${\mathrm{ MeV}}/c^2$\fi}}
\new{\GeV}  {{\ifmmode {\mathrm{ GeV}}\else ${\mathrm{ GeV}}$\fi}}
\new{\GeVc} {{\ifmmode {\mathrm{ GeV}}/c\else ${\mathrm{GeV}}/c$\fi}}
\new{\GeVcc}{{\ifmmode {\mathrm{ GeV}}/c^2\else ${\mathrm{GeV}}/c^2$\fi}}
\new{\TeV}  {{\ifmmode {\mathrm{ TeV}}\else ${\mathrm{ TeV}}$\fi}}

\new{\Mh}     {{\ifmmode M_{\mathrm{ H}}
                \else $M_{\mathrm{H}}$\fi}}
\new{\Mz}     {{\ifmmode M_{\mathrm{Z}}
                \else $M_{\mathrm{Z}}$\fi}}
\new{\Mzsq}   {{\ifmmode M^2_{\mathrm{ Z}}
                \else $M^2_{\mathrm{Z}}$\fi}}
\new{\as}[1]  {{\ifmmode\alpha^{#1}_s
                \else$\alpha^{#1}_s$\fi}}
\new{\asx}[1]  {{\ifmmode a^{#1}_s
                \else $a^{#1}_s$\fi}}
\new{\asb}[1] {{\ifmmode\overline{\alpha}^{#1}_s
                \else $\overline{\alpha}^{#1}_s$\fi}}
\new{\asmz}   {{\ifmmode\alpha_s(\Mzsq)
                \else $\alpha_s(\Mzsq)$\fi}}
\new{\lqcd}   {{\ifmmode\Lambda_{\mathrm{ QCD}}
                \else $\Lambda_{\mathrm{ QCD}}$\fi}}

\def\Gosam{{{\sc GoSam}}}
\def\GOSAM{{{\sc GoSam}}}
\def\gosam{{{\sc GoSam}}}

\def\SAMURAI{{{\sc Samurai}}}
\def\Sherpa{{{\sc Sherpa}}}

\def\C++{{{\sc c++}}}
\def\Powheg{{{\sc Powheg}}}
\def\herwig{{{\sc Herwig++}}}
\def\Pythia{{{\sc Pythia}}}

\def\Golem{{{\sc Golem95C}}}
\def\GOLEMVC{{{\sc Golem95C}}}

\def\fastjet{{{\sc FastJet}}}
\def\Ninja{{{\sc Ninja}}}

\def\MadLoop{{{\sc MadLoop}}}
\def\madloop{{{\sc MadLoop}}}
\def\Madspin{{{\sc MadSpin}}}
\def\amcnlomglong{{{\sc MadGraph5\_aMC@NLO}}}

\def\amc{{{\sc MG5\_aMC}}}
\def\amcnlomg{{{\sc MG5\_aMC}}}
\def\mgamc{{{\sc MG5\_aMC}}}

\def\ttH{$t\bar{t}H$}
\def\ttyy{$t\bar{t}\gamma\gamma$}

\journalname{Eur. Phys. J. C}
\begin{document}

\title{Spin Polarisation of $t\bar{t}\gamma\gamma$ production at NLO+PS with \Gosam{} interfaced to \amcnlomglong{}
}


\author{Hans~van~Deurzen\thanksref{e1,addr1}
        \and
        Rikkert~Frederix\thanksref{e2,addr2} 
        \and
        Valentin~Hirschi\thanksref{e3,addr3}
        \and
        Gionata~Luisoni\thanksref{e4,addr1}
        \and
        Pierpaolo Mastrolia\thanksref{e5,addr1,addr4}
        \and
        Giovanni Ossola\thanksref{e6,addr5,addr6}
}

\thankstext{e1}{hdeurzen@mpp.mpg.de}
\thankstext{e2}{rikkert.frederix@tum.de}
\thankstext{e3}{vahirsch@slac.stanford.edu}
\thankstext{e4}{luisonig@mpp.mpg.de}
\thankstext{e5}{pierpaolo.mastrolia@cern.ch}
\thankstext{e6}{gossola@citytech.cuny.edu}


\institute{Max-Planck-Institut f\"ur Physik, 
  F\"ohringer Ring 6, 80805 M\"unchen, Germany \label{addr1}
           \and
      Physik Department T31, Technische Universit\"at
  M\"unchen, James-Franck-Str.~1, 85748 Garching, Germany \label{addr2}
           \and
           SLAC, National Accelerator Laboratory, 
  2575 Sand Hill Road, Menlo Park, CA 94025-7090, USA \label{addr3}
           \and
                Dipartimento di Fisica e Astronomia,  Universit\`a di Padova and INFN Sezione di Padova, via Marzolo 8,  35131 Padova, Italy \label{addr4}
           \and
           New York City College of Technology, 
  City University of New York, 300 Jay Street, Brooklyn NY 11201, USA \label{addr5}
           \and
           The Graduate School and University Center, 
  City University of New York, 365 Fifth Avenue, New York, NY 10016, USA \label{addr6}
}

\date{Received: date / Accepted: date}

\maketitle

\begin{abstract} 
 We present an interface between the multipurpose Monte
  Carlo tool \amcnlomglong{} and the automated amplitude
  generator \Gosam{}. As a first application of this novel framework, we
  compute the NLO corrections to $pp \to$ \ttH{} and $pp
  \to$ \ttyy{} matched to a parton shower. In the phenomenological
  analyses of these processes, we focus our attention on observables
  which are sensitive to the polarisation of the top quarks.
\end{abstract}

\section{Introduction}
\label{Sec:intro}
The development of automated tools for precise calculations of total
cross sections and differential distributions in high-energy
collisions has undergone a dramatic acceleration in the last decade.
While Leading-Order (LO) tools, based on automated tree-level
calculations, have been available for a long
time~\cite{Stelzer:1994ta, Kanaki:2000ey, Mangano:2002ea,
Corcella:2002jc, Maltoni:2002qb, Gleisberg:2003xi,
Boos:2004kh,Pukhov:2004ca, Alwall:2007st, Kilian:2007gr,
Cafarella:2007pc}, the needs of the experimental analyses at the Large
Hadron Collider (LHC) and a deeper understanding of the structure of
scattering amplitudes~\cite{Bern:2007dw, Britto:2010xq, Ellis:2011cr}
at one loop led to the development of several computer frameworks for
the automated computation of loop matrix
elements~\cite{Giele:2008bc,Berger:2008sj,Cullen:2011ac,Cascioli:2011va,Actis:2012qn,Cullen:2014yla}
and physical observables at Next-to-leading-Order (NLO)
accuracy~\cite{Gleisberg:2008ta,Bevilacqua:2011xh,Hirschi:2011pa,Badger:2012pg,Alioli:2010xd,Bellm:2013lba,
Alwall:2014hca}.  Moreover, techniques to properly deal with the
merging of different multiplicities in the final states and the
matching~\cite{Frixione:2002ik,Nason:2004rx}~to parton shower were
successfully developed~\cite{Hamilton:2012np,
Hoeche:2012yf,Gehrmann:2012yg,Frederix:2012ps,Lonnblad:2012ix} and are
nowadays available.

Advanced automated calculations have been performed by embedding the
codes for generating virtual corrections at NLO precision, the so
called One Loop Providers (OLPs), within a Monte Carlo program
(MC). The interplay between MCs programs and OLPs is controlled by
means of interfaces, which allow the user to get direct access to the
main features of the MC code, bypassing the need of knowing the
technical details of the OLP, which is ideally an inner engine within
the MC generator.  Many of such interfaces are based on the standards
settled by the Binoth Les Houches Accord
(BLHA)~\cite{Binoth:2010xt,Alioli:2013nda}, which defines
specifications of the communication between MCs and OLPs.

The LHC has recently started Run II, collecting data at an energy
scale never explored before. Within this activities automated
multi-purpose tools for particle collisions simulation are of
fundamental importance for comparing theoretical predictions with
experimental data, thereby extracting important information about the
Standard Model (SM) and exploring all traces of deviations from it.
The need for flexible tools which at the same time can provide
accurate predictions, both in the SM and Beyond (BSM), may become of
primary relevance in the near future. It is therefore important to be
able to connect different tools to increase the reliablility of
results.

With this goals in mind, we present the interface between the
multipurpose NLO Monte Carlo tool \amcnlomglong~(\amcnlomg)~\cite{Alwall:2014hca}~and
the automated one-loop amplitude
generator \Gosam~\cite{Cullen:2014yla}.
The advantage of this combination is twofold.
On the one hand, this tandem allows the user of \amc{} to switch
between two options of OLPs, namely between the inhouse code {\sc
MadLoop} \cite{Hirschi:2011pa} fully integrated directly in the MC
distribution package, and \Gosam{}. Thus, the user can experience the
evaluation of NLO virtual corrections by means of two alternative
solutions corresponding to different algorithms and methods of
generation and evaluation of Feynman amplitudes.  On the other
hand, \Gosam{} is interfaced to several MCs codes, like
\Sherpa \cite{Gleisberg:2008ta}, \herwig \cite{Bellm:2013lba}, \Powheg \cite{Alioli:2010xd}, beside \amcnlomg{}, therefore the user of the MCs can explore and compare the different features of the event generators, without being biased by the performances of the OLPs, since they all can be run using \Gosam{}.

As an illustration of the novel framework \amcnlomg{} + \Gosam{}, we
present its application to the NLO corrections to $pp \to$ \ttH,
$H\to\gamma\gamma$ and the continuum $pp \to$ \ttyy{} matched to a
parton shower. The production of a Higgs Boson in association with a
pair of top anti-top quarks is an important process to directly study
the Yukawa coupling of the Higgs Boson with massive fermions. Such a
channel, and its corresponding backgrounds, were recently the subject
of detailed studies, both at LO~\cite{Biswas:2014hwa,
Denner:2014wka,Santos:2015dja} and NLO~\cite{Kardos:2014pba}
precision. Very recently, new analyses have appeared which further
extend these studies including the decay of the top and anti-top quark
into bottom quarks and leptons~\cite{Denner:2015yca}, considering the
production of a top-quark pair in conjunction with up to two vector
bosons~\cite{Maltoni:2015ena}, and exploring the CP-structure of the
top-Higgs coupling~\cite{Buckley:2015vsa}.  In the phenomenological
analysis contained in this paper, we focus our attention on
observables sensible to the polarization of the top quarks, such as
angular variables which involve the decay of the top quark.

The paper is organized as follows: in Section~\ref{Sec:setup}, after a
general introduction to the \Gosam{} and \amcnlomg{} codes, we will discuss
the interface between the two frameworks and its validation. In
Section~\ref{Sec:results}, we will present an application of
the \gosam + \amcnlomg{} interface, namely the study of NLO
corrections to $pp \to$ \ttH, $H\to\gamma\gamma$ and $pp \to$ \ttyy{}
matched to a parton shower. Finally in Section~\ref{Sec:conclusions},
we will draw our conclusions.

\section{Computational setup}
\label{Sec:setup}
For the computations contained in this paper, the automated one-loop
amplitude generator \Gosam{} has been fully interfaced to the
\amcnlomg{}  Monte Carlo framework.

In this section, we briefly review the main characteristics of each of
these tools and describe the details of the interface between them,
which allows to use one-loop amplitudes generated by \Gosam{} within
\amcnlomg{}. Finally we discuss the validation of the interface by
means of a comparison with an independent framework.

\subsection{\Gosam{}}
\label{Sec:gosam}
The main idea that distinguishes the \gosam{} framework~\cite{Cullen:2014yla} from other
codes for the automated generation of one-loop amplitudes is the combined use of
automated diagrammatic generation and algebraic manipulation in $d = 4
-2 \epsilon$ dimensions, thus providing analytic expressions for the
integrands, with $d$-dimensional integrand-level reduction techniques,
or tensorial reduction.  Amplitudes are automatically generated via
Feynman diagrams and, according to the reduction algorithm selected by
the user, are algebraically manipulated and cast in the most
appropriate output~\cite{Nejad:2013ina,Nogueira:1991ex, Vermaseren:2000nd,
Reiter:2009ts, Cullen:2010jv, Kuipers:2012rf}. The individual program
tasks are controlled by means of a python code, while the user only
needs to prepare an input card to specify the details of the process
to be calculated without worrying about internal details of the code
generation.

After the generation of all contributing diagrams, the virtual
corrections are evaluated using the integrand reduction via Laurent
expansion~\cite{Mastrolia:2012bu}, provided by
{\Ninja}~\cite{vanDeurzen:2013saa, Peraro:2014cba}, or the
$d$-dimensional integrand-level reduction
method~\cite{Ossola:2006us,Ellis:2008ir,Mastrolia:2008jb}, as
implemented in \SAMURAI~\cite{Mastrolia:2010nb,vanDeurzen:2013pja}. Alternatively,
the tensorial decomposition provided by
{\Golem}~\cite{Binoth:2008uq,Heinrich:2010ax,Cullen:2011kv,Guillet:2013msa} is also
available.  The scalar loop integrals can be evaluated using \GOLEMVC,  {\sc
OneLOop}~\cite{vanHameren:2010cp}, or {\sc
QCDLoop}~\cite{vanOldenborgh:1990yc,Ellis:2007qk}.

The \gosam\ framework can be used to generate and evaluate one-loop corrections
in both QCD and electroweak theory~\cite{Chiesa:2015mya}. Model files for Beyond Standard
Model (BSM) applications generated from a Universal FeynRules Output
(\texttt{UFO})~\cite{Degrande:2011ua, Alloul:2013bka} or
with \texttt{LanHEP}~\cite{Semenov:2014rea} are also supported.  A
model file which contains the effective Higgs-gluon couplings that
arise in the infinite top-mass limit is also available in the current
distribution and it was successfully used to compute the virtual
corrections for the production of a Higgs boson in association with 2
and 3 jets~\cite{vanDeurzen:2013rv,Cullen:2013saa}.

The computation of physical observables at NLO accuracy, such as cross
sections and differential distributions, requires to combine the
one-loop results for the virtual amplitudes obtained with \gosam, with
other parts of the calculations, namely the computation of the real
emission contributions and of the subtraction terms, needed to control
the cancellation of IR singularities. In some of the earlier
calculations performed with \gosam~\cite{Greiner:2012im,
Cullen:2012eh, Gehrmann:2013aga,Gehrmann:2013bga,Greiner:2013gca}, the
problem was solved by means of an ad hoc adaptation of the {\sc
MadDipole-Madgraph4-MadEvent} framework~\cite{Stelzer:1994ta,Frederix:2008hu,Frederix:2010cj,Alwall:2007st}.

Far more efficiently, this task can be performed by embedding the
calculation of virtual corrections within a multipurpose Monte Carlo program (MC),
that can also provide the phase-space integration, which is what is pursued in this work.  In this case, the
MC takes control over the different stages of the calculation, in
particular the phase space integration and the event generation, and
calls \Gosam{} at runtime to obtain the corresponding value of the
one-loop amplitude at the given phase space points.  This approach has
the great advantage of making available to the user all the advanced
features that the MC generator provides, for example to allow for
parton showering, further decays of the final-state hard particles
retaining spin correlation and merging of different final state
multiplicities.

While in the present paper we will focus on the interface
between \gosam\ and \mgamc, it is worth mentioning that a number of
phenomenological results can be found in the literature which were
obtained by combining \Gosam{} with other MC programs, in particular
with Sherpa~\cite{Gleisberg:2008ta,Hoeche:2013mua,vanDeurzen:2013xla,Heinrich:2013qaa,Greiner:2015jha}, \herwig~\cite{Bellm:2013lba, Butterworth:2014efa}, and \Powheg~\cite{Alioli:2010xd,Luisoni:2013cuh, Luisoni:2015mpa}.

\subsection{\amcnlomglong{}}
\label{Sec:amcnlo}
The \amcnlomg\ framework~\cite{Alwall:2014hca}~has been developed to
be able to generate events and compute differential cross sections
with a high-level of automation. The central idea behind the code is
that the structure of cross sections is essentially independent of the
process under consideration. Therefore, once this structure has been
set up, any cross section can be computed within the framework. For
example, even though the matrix elements are process and theory dependent,
they can be computed from a very limited set of instructions based on
the Feynman rules.

The core of the \amcnlomg\ framework is based on tree-level amplitude
generation, since in this code the matrix elements used in both LO and
NLO computations are constructed from tree-level Feynman diagrams. The
generation of these amplitudes is based on three elements which are
key to taming the complexity of the computation as the number of
external particles increases: colour decomposition, helicity
amplitudes and recycling of identical substructures between
diagrams. The internal algorithms used have been described extensively
in Ref.~\cite{Alwall:2011uj} and
Ref.~\cite{Hirschi:2011pa, Ossola:2006us, Ossola:2007ax, Ossola:2008xq, Cascioli:2011va,Alwall:2014hca}
for the generation of tree-level and one-loop matrix elements,
respectively.

Beyond the lowest order in perturbation theory, intermediate
contributions to the computation of (differential) cross sections are
plagued by divergences. In particular the soft/collinear divergences
in the numerical phase-space integration over the real-emission
(Bremstrahlung) corrections are non-trivial to deal with. 
In \amcnlomg, the FKS subtraction
method \cite{Frixione:1995ms, Frederix:2009yq} is used to factor out
the singularities in order to cancel them analytically with
singularities present in the virtual corrections. The FKS subtraction
is based on partitioning the phase-space, in which each phase-space
region has at most one collinear and one soft singularity. These
singularities are subtracted before performing the numerical
phase-space integration by using a straight-forward
plus-description. The subtraction terms have been integrated
analytically, using dimensional regularisation, once and for all,
resulting in explicit poles in $1/\epsilon$ and $1/\epsilon^2$, which
cancel similar poles in the virtual corrections and the PDFs.

Independently of which code one uses for the computation of the
virtual corrections, for example \MadLoop~\cite{Hirschi:2011pa} or
\Gosam~\cite{Cullen:2014yla}, optimisations in the phase-space
integration of these contributions are used, as described in
Ref.~\cite{Alwall:2014hca}. That is, during the phase-space
integration an approximation of the virtual corrections based on the
born matrix elements is created dynamically. These approximate virtual
matrix elements are very fast to evaluate and can therefore be
efficiently integrated numerically with high statistics. The small
difference between the approximate and the exact virtual corrections
is relatively slow to evaluate for a given phase-space point, but
given that this is a very small contribution to the final result, the
requirement on the relative precision with which it needs to be computed can be
relaxed. Therefore, using low statistics for this complicated
contribution suffices, greatly reducing the overall computational
time.

To match the short-distance matrix elements to a parton shower, the
 framework of \amcnlomg\ employs the MC@NLO
technique~\cite{Frixione:2002ik} available for Pythia~6~\cite{Sjostrand:2006za}, Herwig~6~\cite{Corcella:2000bw}, Pythia~8~\cite{Sjostrand:2014zea}~and Herwig++~\cite{Bahr:2008pv}. In this
method, the possible double counting between the NLO corrections and
the parton shower is accounted for by explicitly subtracting the
parton shower approximation for the emission of a hard parton from the
real-emission contributions, and the parton shower approximation of a
non-emitting parton from the virtual corrections. To consistently merge
various multiplicities at NLO accuracy and match them to the parton shower, 
the FxFx merging method~\cite{Frederix:2012ps}~is available.

\subsection{Interface}
\label{Sec:interface}
The interface between \Gosam{} and \amcnlomg{} is based on the
standards of the first BLHA defined in~\cite{Binoth:2010xt}.
When running the \mgamc{} interactive session, the command
\begin{center}
\texttt{\$ set OLP GoSam}
\end{center}
changes the employed OLP from its default \MadLoop{} to \Gosam{}. 
Alternatively, the file
\begin{center}
\texttt{input/mg5\_configuration.txt}
\end{center}
can be directly edited to include the line \texttt{OLP = GoSam}.

The BLHA order and contract file system allows for a basic
communication between the two codes to exchange the most fundamental information
about the number and type of subprocesses, the powers of the couplings
involved in the specific process, the schemes in which the computation
should be performed and also the value of parameters like masses and
widths. For static parameters, which do not change at each phase space
point, but stay constant during the MC integration and event
generation, the information is passed via a SUSY Les Houches Accord
(SLHA) parameter file. This is created by \amcnlomg{} and read by
\Gosam. The path to this SLHA file is specified in the order file with the
key word \texttt{ModelFile}. An example generated for the computation
of \ttyy{} is shown in Figure~\ref{fig:order_and_contract}. For
parameters which instead may change at each phase space point, the
BLHA 1 interface defines an array to pass the numerical value of
dynamical variables.  The definition and order of the parameters
passed through this array is set in the order file using the keyword
\texttt{Parameters}.  Although in principle extendible to up to ten
parameters, at present only the first entry is used, to communicate
the value of $\alpha_S$.

\begin{figure}[h]
\begin{minipage}[t]{.49\textwidth}
\footnotesize{
\begin{Verbatim}[frame=single]
#OLE_order written by MadGraph5_aMC@NLO

MatrixElementSquareType CHaveraged
CorrectionType          QCD
IRregularisation        CDR
AlphasPower             2
AlphaPower              2
NJetSymmetrizeFinal     Yes
ModelFile               ./param_card.dat
Parameters              alpha_s


# process
21 21 -> 22 22 6 -6
2 -2 -> 22 22 6 -6
1 -1 -> 22 22 6 -6
-2 2 -> 22 22 6 -6
-1 1 -> 22 22 6 -6
\end{Verbatim}
}
\end{minipage}
\hfill
\noindent
\begin{minipage}[t]{.49\textwidth}
\raggedleft\footnotesize{
\begin{Verbatim}[frame=single]
# vim: syntax=olp
#@OLP GoSam 2.0.0
#@IgnoreUnknown True
#@IgnoreCase False
#@SyntaxExtensions
MatrixElementSquareType CHaveraged | OK
CorrectionType QCD | OK
IRregularisation CDR | OK
AlphasPower 2 | OK
AlphaPower 2 | OK
NJetSymmetrizeFinal Yes | OK #Ignored by OLP
ModelFile ./param_card.dat | OK
Parameters alpha_s | OK
21 21 -> 22 22 6 -6 | 1 2
2 -2 -> 22 22 6 -6 | 1 0
1 -1 -> 22 22 6 -6 | 1 3
-2 2 -> 22 22 6 -6 | 1 1
-1 1 -> 22 22 6 -6 | 1 4
\end{Verbatim}
}
\end{minipage}
\caption{Example of order file (above) and contract file (below).}
\label{fig:order_and_contract}
\end{figure}

Further customization of the one-loop amplitudes which need to be
generated can be achieved by editing a separate input file for
\Gosam. In the file \texttt{gosam.rc}, additional information can be specified, i.e. the
model, the particle content of the loop diagrams, and the number of
active flavours. Here it is also possible to define
ad hoc filters to remove unwanted diagrams or loop contributions which
are known to be negligible or vanishing, which may however, if kept in the calculation, introduce
numerical instabilities or slow down the evaluation. Some
settings are present by default, but many more can be introduced by the user. We
refer to the Appendix for a more extensive list of possible
options. The \Gosam{} input file needs to be edited by hand and can be
found at
 \begin{center}
\texttt{\small Template/loop\_material/OLP\_specifics/GoSam/gosam.rc},
\end{center} 
in the \amcnlomg{} repository, or at
\begin{center}
\texttt{OLP\_virtuals/gosam.rc},
\end{center}
in the folder that is generated by \amcnlomg{} automatically when a
new process is started. After the input file is ready, any NLO process
can be generated following the general \amcnlomg{} procedure.

The interface between \amcnlomg{} and \Gosam{} is available starting
from \amcnlomg{} version 2.3.2.2. The two codes can be downloaded from the following URL:
\begin{itemize}
\item[]{\small \amcnlomg: \texttt{http://amcatnlo.web.cern.ch/amcatnlo/}}
\item[]{\small \Gosam: \texttt{http://gosam.hepforge.org/}}
\end{itemize}

\subsection{NLO predictions and validation}
\label{Sec:validation}

To validate the interface, and consequently the results we present in
Section~\ref{Sec:results}, several cross checks were performed.  The
loop amplitudes of \gosam{} and \madloop{} were compared for
single phase space points and also at the level of the total cross
section for a number of different processes, as presented in a
dedicated table in~\cite{hansthesis}. Furthermore, for $pp\rightarrow
t\bar{t}\gamma\gamma$, a fully independent check was also performed by
computing the same cross section using \gosam{} interfaced
to \Sherpa{}. The comparison between the results obtained
with \mgamc{}+\gosam{}, \Sherpa{}+\gosam{}, and \mgamc{}+\madloop{} is
presented Table~\ref{table:xstable8}, where
we report the total integrated cross sections for LO and NLO at a
center-of-mass energy $\sqrt{s}=8$ TeV. 

\begin{table*}[b!] 
\begin{center}
\begin{tabular}{ c c c c } \toprule
\phantom{\Big{|}}$\sigma_{t\bar{t}\gamma\gamma}$, \scriptsize{$\sqrt{s} = 8$ TeV} & \scriptsize{{\sc MG5\_aMC} + \madloop{}} & \scriptsize{{\sc MG5\_aMC} + \gosam{}} & \scriptsize{\Sherpa{}+\gosam{}} \\ 
\midrule
\phantom{\Big{|}} LO [pb]   & \multicolumn{2}{c}{$1.0241\pm 5.50\cdot 10^{-4}$} & $1.0246\pm 3.51\cdot 10^{-4}$\\
\phantom{\Big{|}} NLO [pb] & $1.3507\pm 5.85\cdot 10^{-3}$ & $1.3432\pm 5.16\cdot 10^{-3}$ & $1.3593\pm 1.80\cdot 10^{-3}$ \\ \bottomrule
\end{tabular}
\caption{Total cross sections in picobarns, at center-of-mass energy $\sqrt{s}=8$ TeV, for combinations of MCs and OLPs, at LO and NLO.}
\label{table:xstable8} 
\end{center}
\end{table*}

\begin{figure}[ht] 
\bcen
{\includegraphics[width=0.49\textwidth]{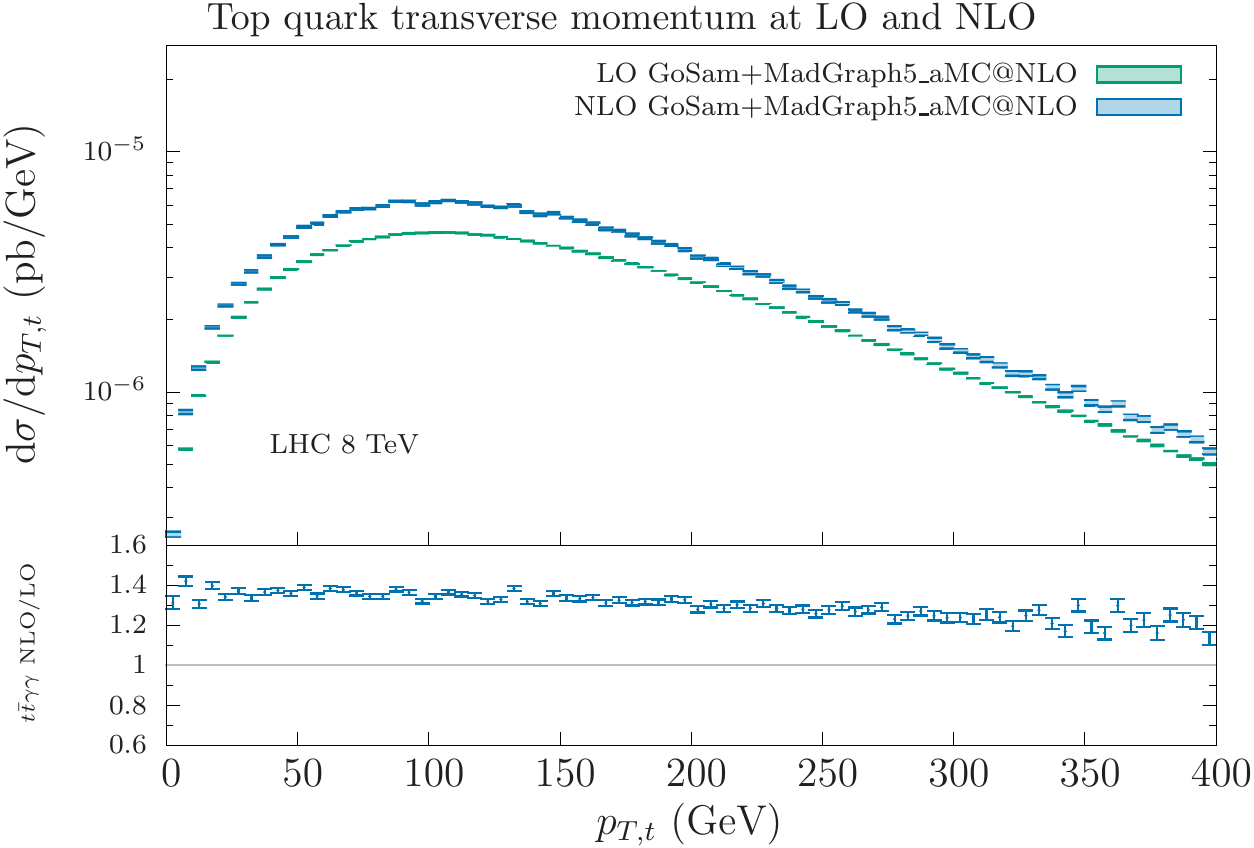}} 
 {\includegraphics[width=0.49\textwidth]{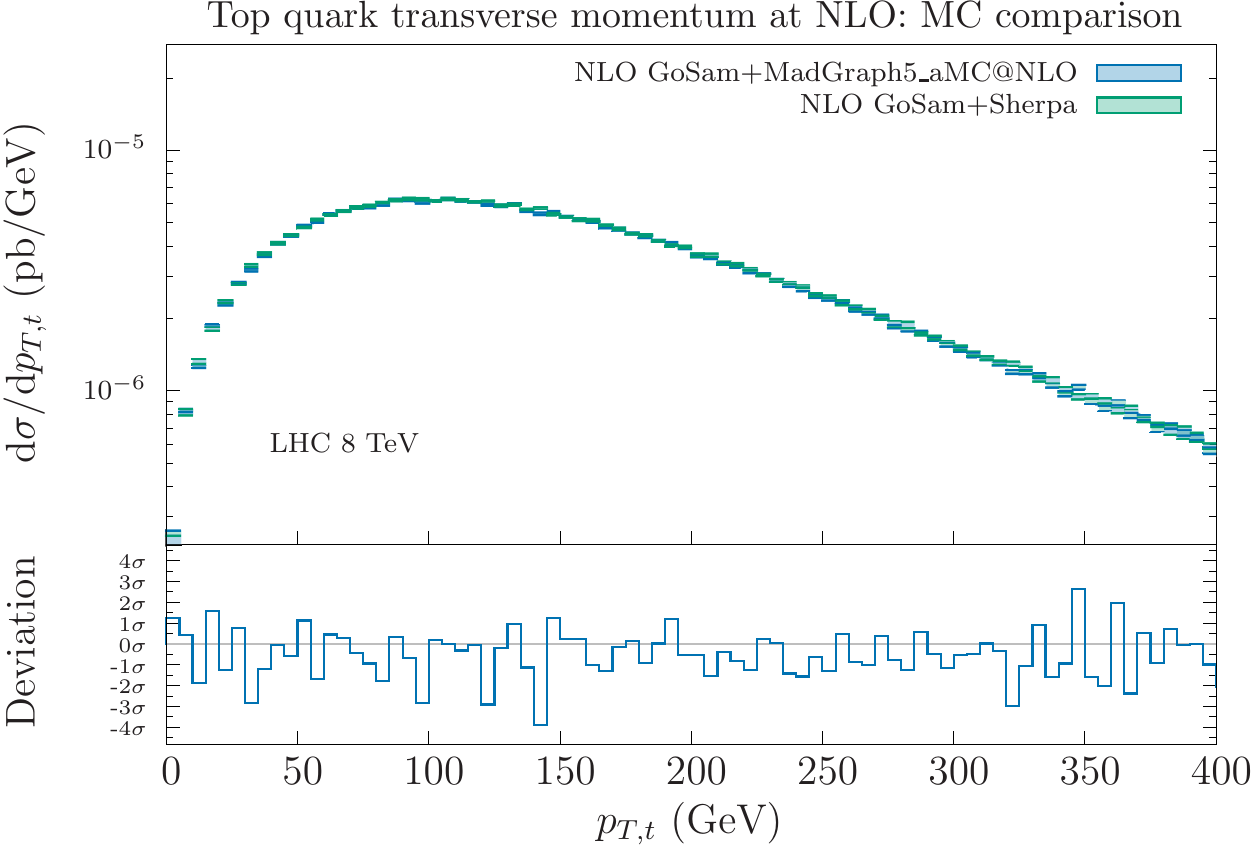}}\\
\ecen
\caption{%
    Transverse momentum of the top quark in $ p p \to
    t \bar{t} \gamma \gamma$ for the LHC at 8 TeV: LO and NLO 
    distributions obtained with \GOSAM{}+\amcnlomg\ (upper plot) and NLO
    comparison between \GOSAM{}+\amcnlomg\ and \GOSAM{}+\Sherpa\ (lower plot).}
\label{fig:ttyy_pt}
\end{figure}

\begin{figure}[ht] 
\bcen
{\includegraphics[width=0.49\textwidth]{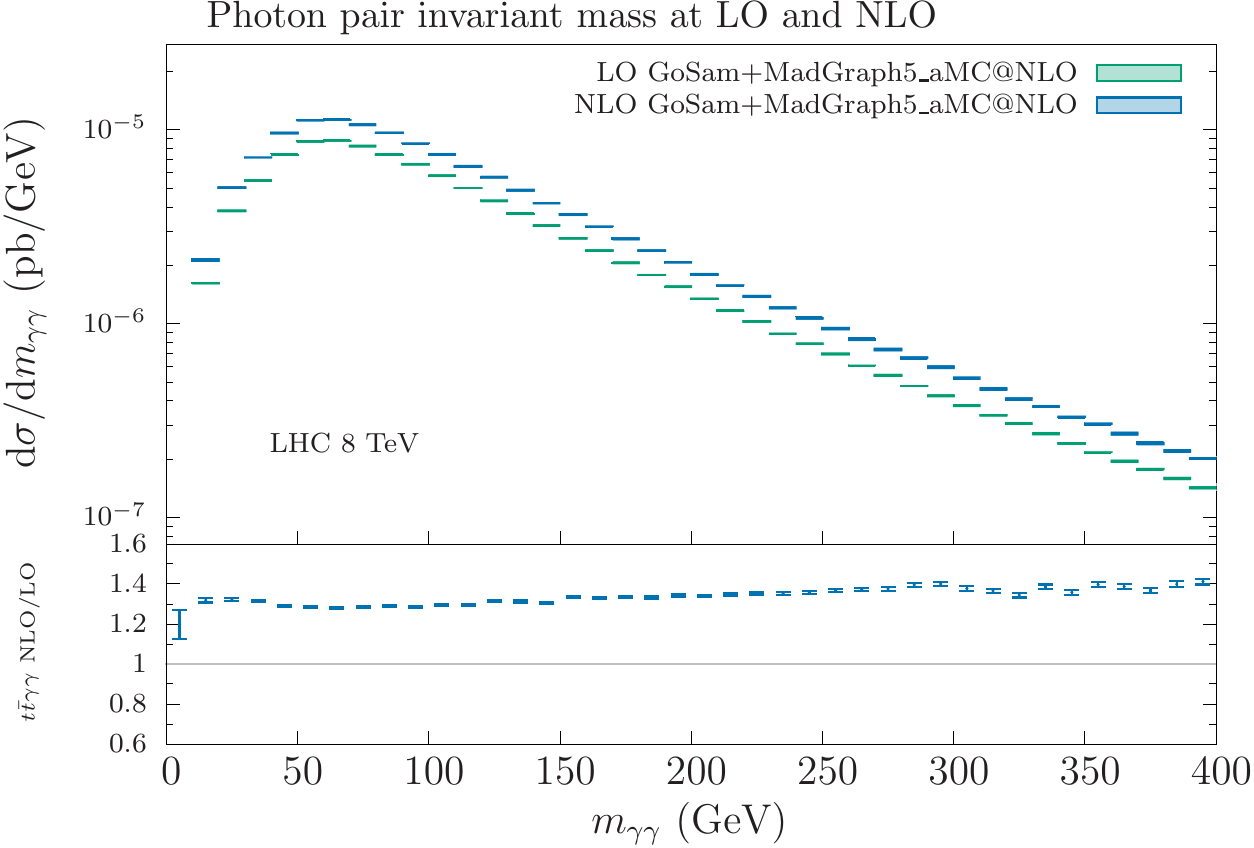}} 
 {\includegraphics[width=0.49\textwidth]{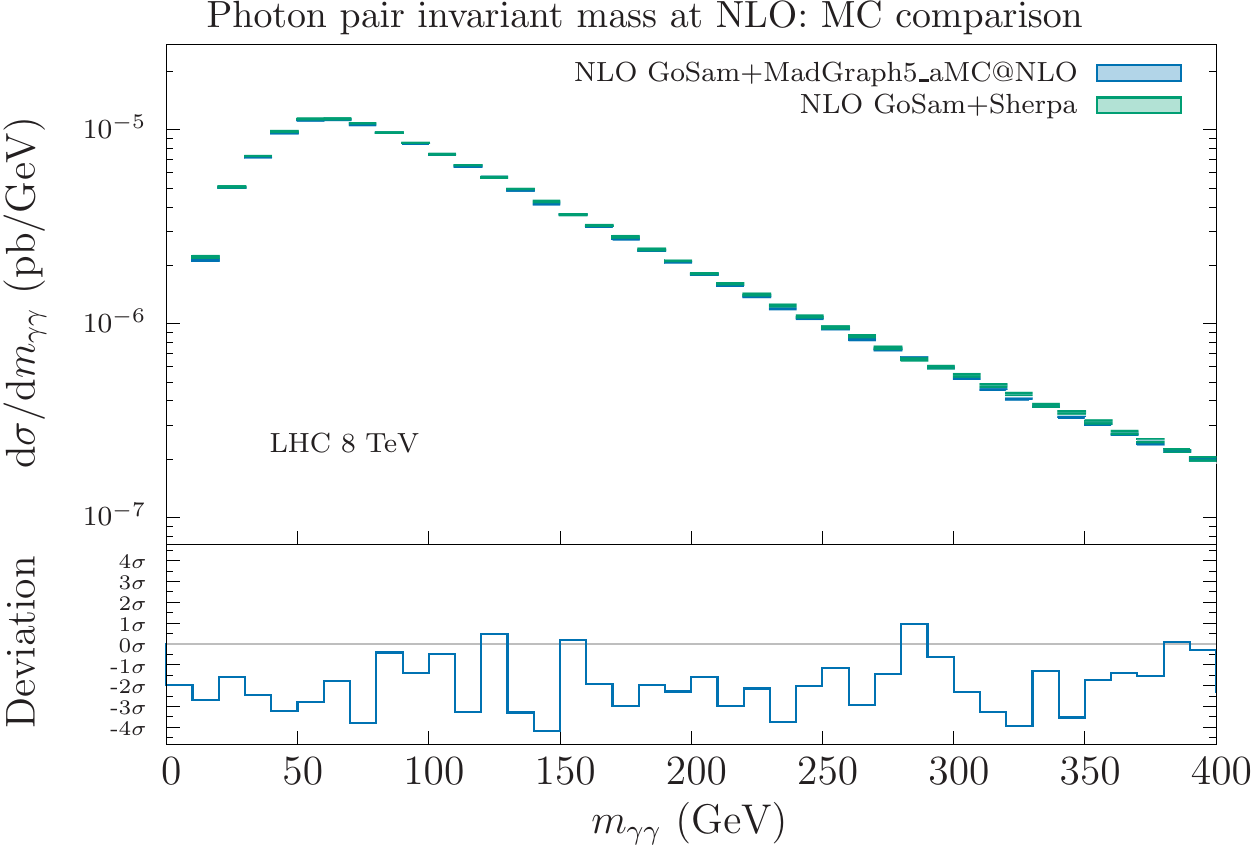}}\\
\ecen
\caption{%
    Photon pair invariant mass distribution in $ p p \to
    t \bar{t} \gamma \gamma$ for the LHC at 8 TeV:
    LO and NLO 
    distributions obtained with \GOSAM{}+\amcnlomg\ (upper plot) and NLO
    comparison between \GOSAM{}+\amcnlomg\ and \GOSAM{}+\Sherpa\ (lower plot).}
\label{fig:ttyy_myy}
\end{figure}

The results shown in this and the following sections are computed using the following
setup. 
The mass of the Higgs was set to $m_H = 125$ GeV, the mass of the top
quark to $m_t= 173.2$ GeV.  We work in the $N_f=5$ model. The value of
the electroweak coupling is set to its low energy limit
$\alpha_{EW}^{-1}=137.0$. The mass of the $Z$ boson was set to $m_Z =
91.1876$ GeV and the value of the Fermi constant to $G_F=1.16639\cdot
10^{-5}$ GeV$^{-2}$, which fixes the electroweak scheme. For the
photons, we used the isolation procedure introduced by
Frixione~\cite{Frixione:1998jh}~with minimal transverse momentum
$p_{\gamma T\text{min}}=20$ GeV, radius of isolation $R_\gamma < 0.4$
and Frixione parameters $n=1.0$ and $\epsilon_\gamma=1.0$.
Furthermore, we applied an isolation radius between the two photons
$R_{\gamma\gamma}=0.4$.  In leading order calculations, we used the
PDF set cteq6L1 \cite{Pumplin:2002vw}. At next-to-leading order, we
instead used the PDF set CT10.  The renormalisation and factorisation
scales are set to $\mu_R = \mu_F = \mu_0$ with
\begin{equation}
\label{eq:scale_ht}
\mu_0 = \frac{\hat{H}_T}{2} = \frac{1}{2}\left(\sum_{\text{final state i}} m_{T,i}\right)
\end{equation}

In Figure~\ref{fig:ttyy_pt}, on the left, we compare LO and NLO
predictions for the transverse momentum of the top quark obtained
with \GOSAM{}+\mgamc{}, while on the right we compare the same NLO
predictions with results obtained using \GOSAM{}+\Sherpa. In Figure~\ref{fig:ttyy_myy} we do the same for the photon pair invariant mass. 
All predictions are computed for a center of mass energy of 8 TeV.

\section{Results}
\label{Sec:results}
In this section we present results at NLO+PS level for the LHC at 13
TeV and compare the \textit{background} process \ttyy, where the
photons are directly radiated from the quarks, with the
\textit{signal} process \ttH{} in which the Higgs boson decays to two
photons. We will refer to the latter simply as ``\ttH''; it should be
understood that we consider only the process with the photonic Higgs
decay \ttH, $H\rightarrow\gamma\gamma$. As a reference, we also include results for the total cross sections and a selection of distributions obtained at 8 TeV.

The study is performed using NLO predictions for \ttH\ and continuum
\ttyy{} production. The top and anti-top quarks are
subsequently decayed semi-leptonically
$t\to W^{+}(\to\,\bar{l}\,\nu_l)b$,
$\bar{t}\to\,W^{-}(\to\,l,\bar{\nu}_l)\bar{b}$ 
with \Madspin~\cite{Frixione:2007zp,Artoisenet:2012st}, taking into
account spin correlation effects, and then showered and hadronised
by means of  \Pythia~8.2, using its default parameters, but with underlying
event turned off. The short-distance events were generated and compared with two
slightly different sets of cuts in order to verify that they had no impact
on the results at the level of the analysis. Apart from
the kinematical requirements on the reconstructed objects, which we will describe below, we use
identical model parameters, renormalisation and factorisation scales,
PDFs and photon isolation as described in Section~\ref{Sec:validation}.

Note that for the background process we neglect effects of photon
bremsstrahlung from the charged top decay products, which can at least
partially be reduced by applying proper kinematical cuts. For the spin
correlation observables, on which we will focus our
attention in the last part of this section, a LO
study~\cite{Biswas:2014hwa} showed that the impact of neglecting these
contributions is present but not dramatic.

The analysis cuts are designed to increase the signal over the
background, but are by no means optimised to maximise the
enhancement. The two photons from the Higgs decay (or the two hard
photons in the \ttyy\ process) are required to be isolated and fulfill
\begin{eqnarray} \label{eq3p1}
p_{T,\gamma}~>~20~\mathrm{GeV},\quad |\eta_{\gamma}|~<~2.5, \\ 
123~\mathrm{GeV}~<~m_{\gamma\gamma}~<~129~\mathrm{GeV}\,,  \nonumber
\end{eqnarray}
where the invariant mass requirement selects a window around the Higgs
boson mass, which reduces the background significantly without
altering the signal strength. Furthermore, we require the events to
have two oppositely charged leptons and two b-jets coming from the top
and anti-top decays. The leptons are selected requiring
\begin{equation}
p_{T,l^{\pm}}~>~10~\mathrm{GeV}, \quad |\eta_{l^{\pm}}|~<~2.7 \,.
\end{equation}
The b-jets are defined to be jets containing at
least one lowest lying B meson. The jets themselves are defined by
clustering all stable hadrons and photons, but excluding the two photons selected using Eq.~(\ref{eq3p1}),
using the anti-$k_T$ algortithm as implemented in the code
\fastjet~\cite{Cacciari:2005hq,Cacciari:2008gp,Cacciari:2011ma}, with
\begin{equation}
\Delta R = 0.4, \quad p_{T,j}~>~20~\mathrm{GeV}, \quad |\eta_{j}|~<~4.7\,.
\end{equation}
We use MC truth information to select the photons from the Higgs decay
(signal) or hard events (background) as well as the leptons and b-jets
coming from the top and anti-top decays. As reconstructed (anti-)top
quark, we use the four-momentum of the (anti-)top quark just before it
decays, as provided in the Pythia~8 event record. In the presence of
these analysis cuts, we obtain the cross-sections reported in
Table~\ref{table:xstableanalysis}.

\begin{table*}[ptb] 
\begin{center}
\begin{tabular}{ c c c }\toprule
\phantom{\Big{|}} \scriptsize{$\sqrt{s} = 8$ TeV} & $pp\rightarrow t\bar{t}H$, $H \rightarrow \gamma\gamma$ & $pp\rightarrow t\bar{t}\gamma\gamma$ \\ 
\midrule
\phantom{\Big{|}} LO [pb] & $2.90(1) \cdot 10^{-7}~{}^{+30\%}_{-21\%}~{}^{+14\%}_{-15\%}$   &   $0.544(1)\cdot 10^{-7}~{}^{+27\%}_{-20\%}~{}^{+14\%}_{-17\%}$ \\
\phantom{\Big{|}} NLO [pb] & $3.71(1) \cdot 10^{-7}~{}^{+4\%}_{-8\%}~{}^{+15\%}_{-16\%}$   &   $0.770(5)\cdot 10^{-7}~{}^{+8\%}_{-9\%}~{}^{+13\%}_{-17\%}$ \\
\phantom{\Big{|}} K-factor & $1.28(1) $ & $1.42(1) $ \\
\midrule
\midrule
\phantom{\Big{|}} \scriptsize{$\sqrt{s} = 13$ TeV} & $pp\rightarrow t\bar{t}H$, $H \rightarrow \gamma\gamma$ & $pp\rightarrow t\bar{t}\gamma\gamma$ \\ 
\midrule
\phantom{\Big{|}} LO [pb] & $8.84(2) \cdot 10^{-7}~{}^{+27\%}_{-20\%}~{}^{+10\%}_{-11\%}$     &       $1.442(2)\cdot 10^{-7}~{}^{+25\%}_{-18\%}~{}^{+10\%}_{-12\%}$ \\
\phantom{\Big{|}} NLO [pb] & $11.77(5) \cdot 10^{-7}~{}^{+6\%}_{-8\%}~{}^{+11\%}_{-12\%}$    &       $2.175(7)\cdot 10^{-7}~{}^{+10\%}_{-10\%}~{}^{+10\%}_{-11\%}$ \\
\phantom{\Big{|}} K-factor & $1.33(1) $ & $1.51(1) $ \\
\midrule
\midrule
NLO Ratio 13TeV/8TeV  & $3.17(2) $ & $ 2.82(2) $ \\
 \bottomrule
\end{tabular}
\caption{Cross sections in picobarns, at a center-of-mass energy of
  $\sqrt{s}=8$ TeV (upper part) and  $\sqrt{s}=13$ TeV (lower part) in the presence of the analysis cuts described
  in the text. The two sets of uncertainties following the cross section correspond to the scale and PDF variations respectively. We also report the ratio between the cross sections at the two center of mass energies.} 
\label{table:xstableanalysis} 
\end{center}
\end{table*}

In the next section we focus our attention on some relevant
observables related to a single particle, whereas in
Section~\ref{Subsec:spinobservables} we will concentrate on
observables which can directly probe correlation effects due to the
top quark polarisation.

In what follows and unless specified otherwise, the plots will always
consist of four distributions. The top curves show the differential
cross sections for a given observable and for both the signal and the
background process. The two middle insets display the relative
uncertainty of the \ttH{} and \ttyy{} predictions respectively. The
scale dependence (transparent band) is estimated by taking the
envelope of the nine predictions obtained by the separate variation of
renormalisation and factorisation scales by factors of 0.5 and 2
around the central scale $\mu_0$ defined in
Equation~(\ref{eq:scale_ht}). For estimating the PDF uncertainty
(dotted lines), we use the Hessian method. Finally, the bottom inset
highlights the differential signal-to-background ratio.

\subsection{Single particle observables}
\label{Subsec:singleobservables}
\begin{figure}[t!]
  \centering
  \includegraphics[width=0.49\textwidth]{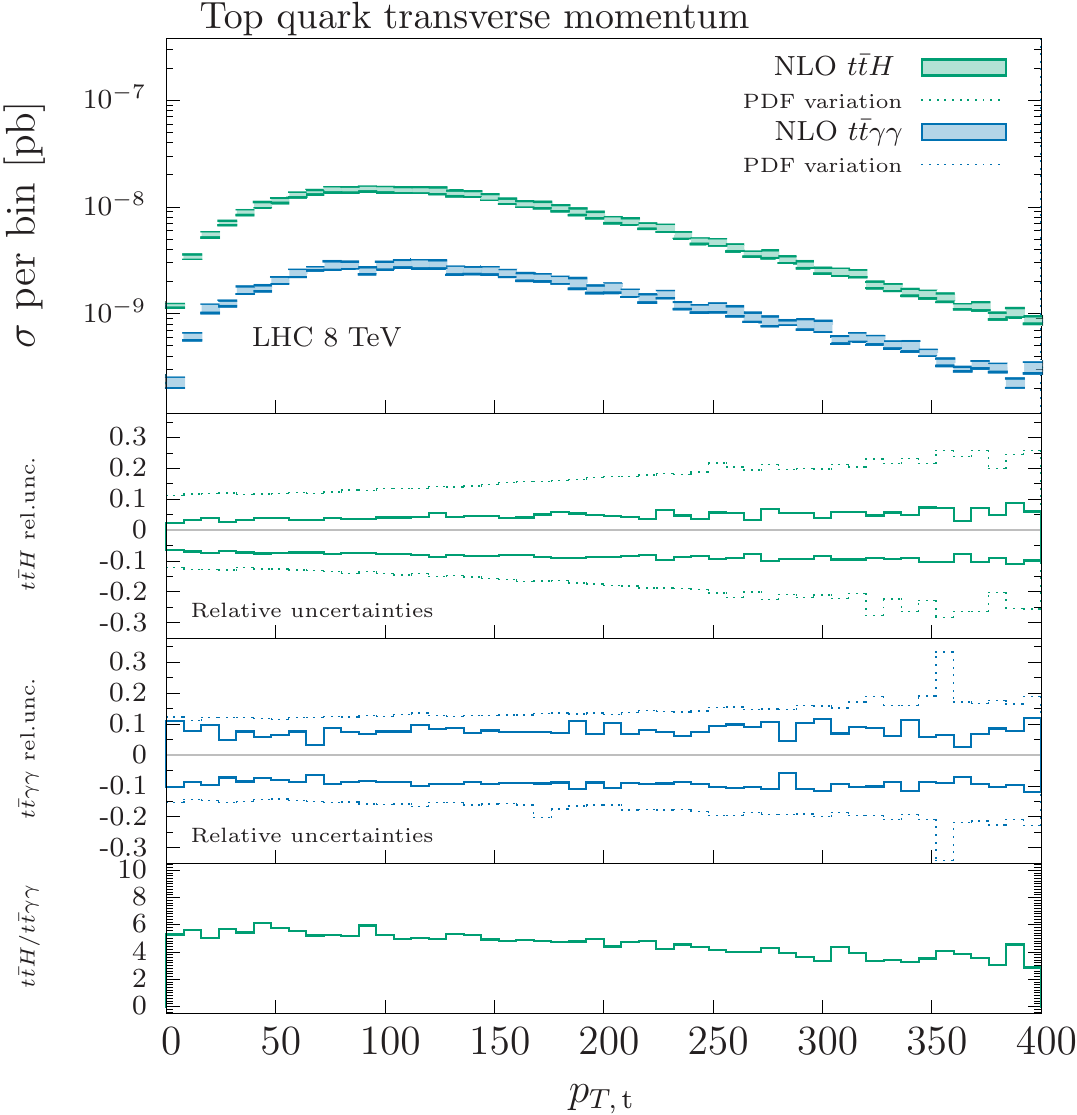}
  \hfill
  \includegraphics[width=0.49\textwidth]{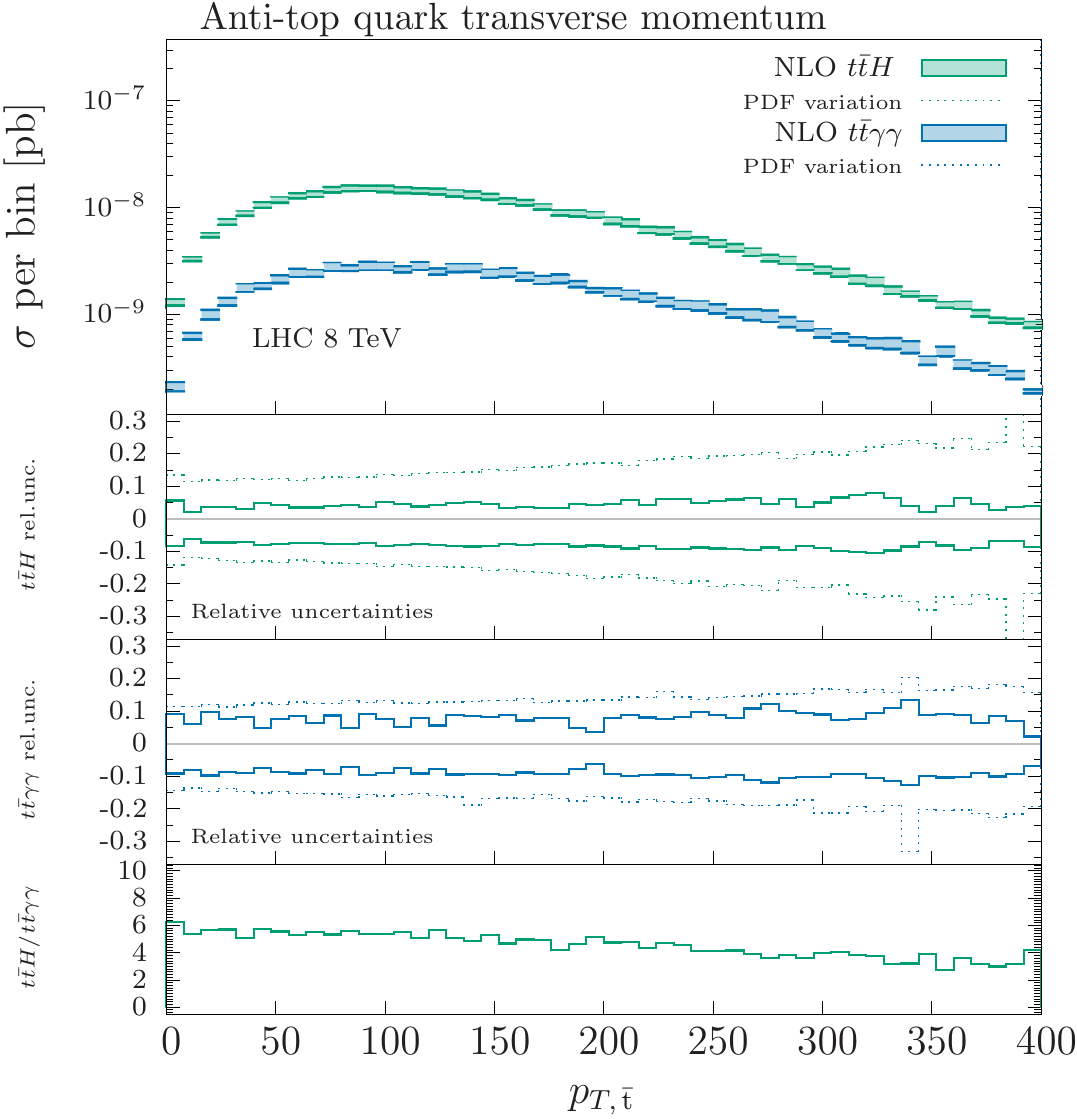}
  \caption{\label{fig:pt_tops8}%
    Transverse momentum distribution of the top quark (above) and
    anti-top quark (below) at 8 TeV.}
\end{figure}

\begin{figure}[t!]
  \centering
  \includegraphics[width=0.49\textwidth]{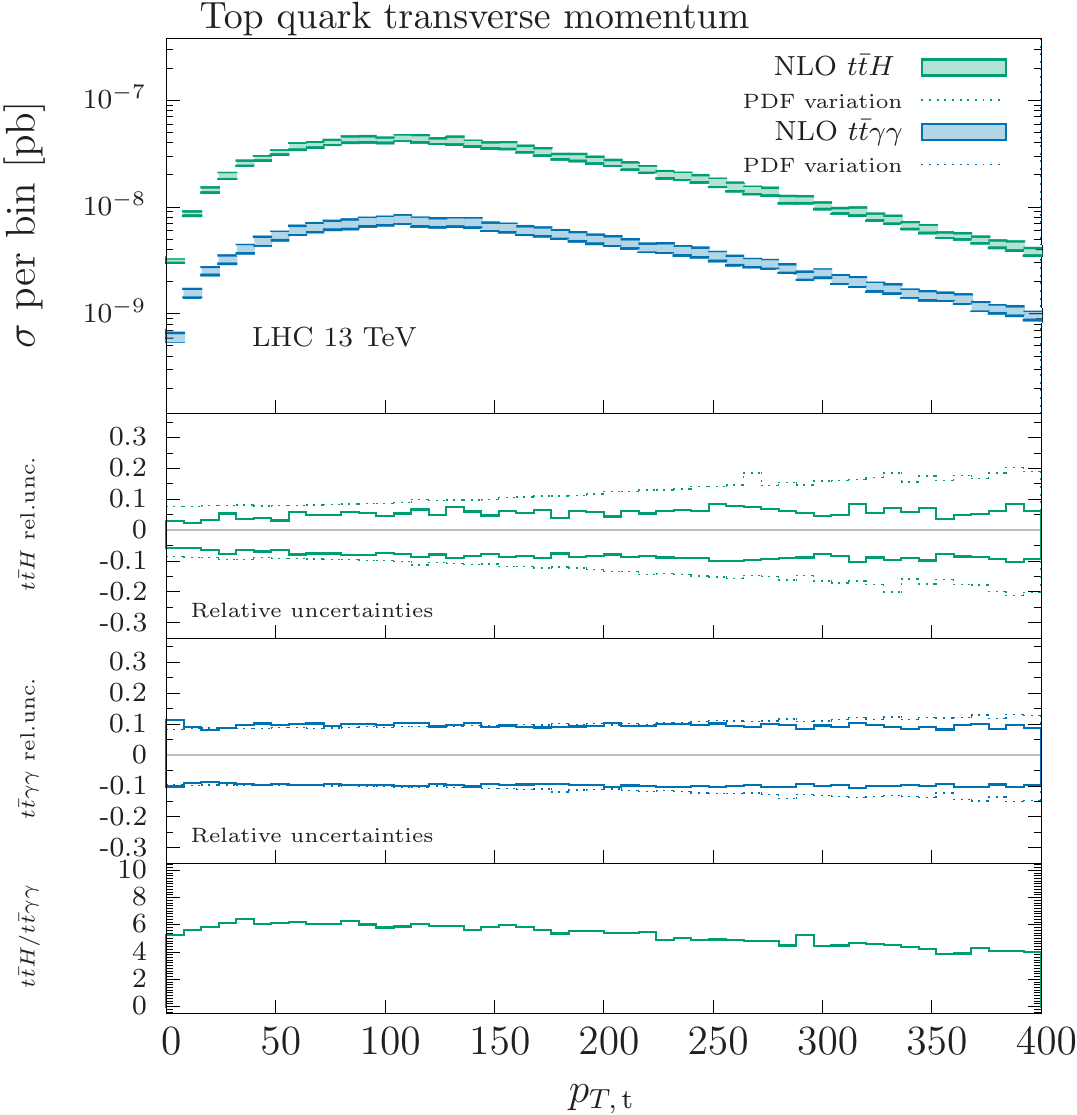}
  \hfill
  \includegraphics[width=0.49\textwidth]{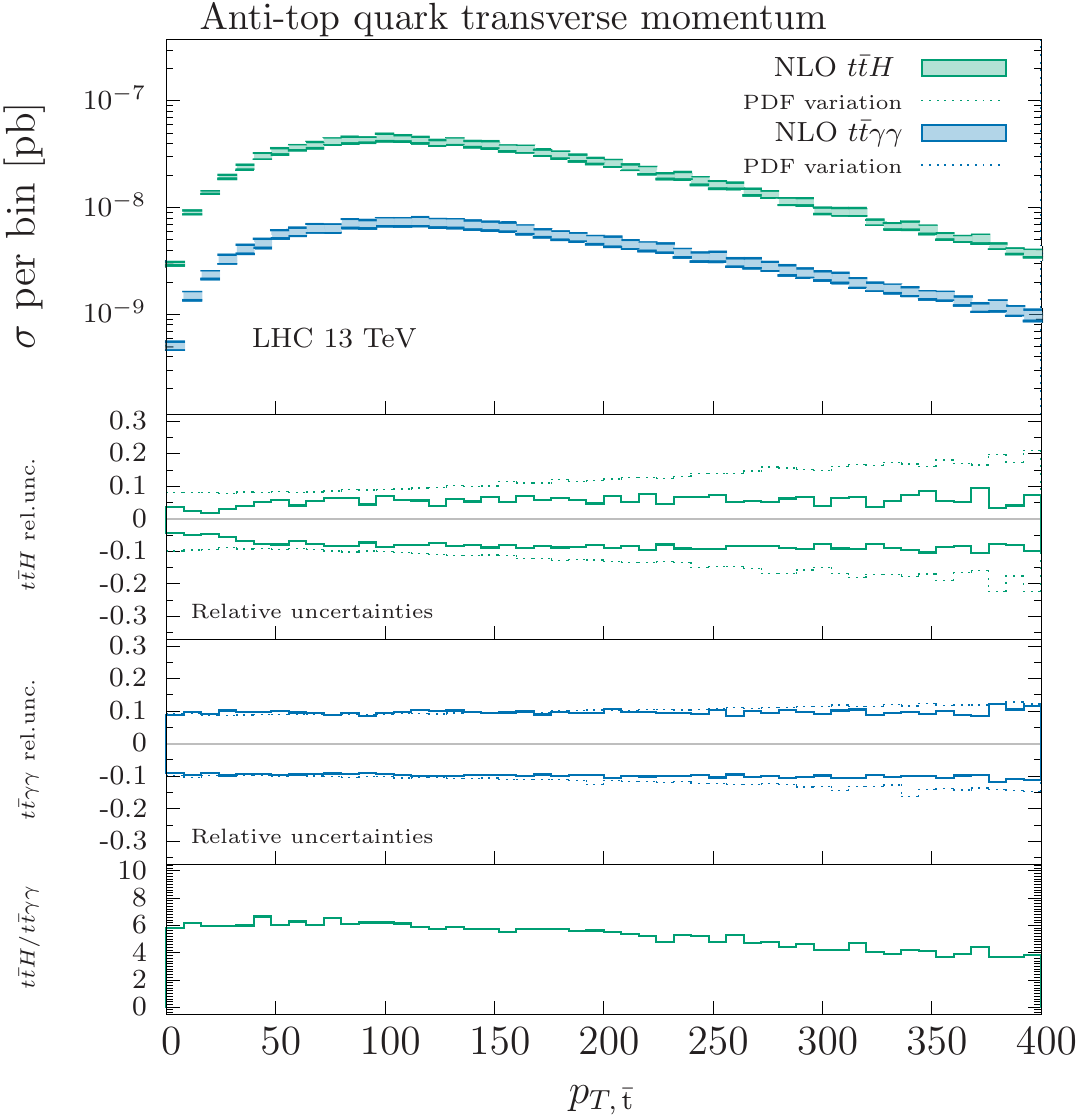}
  \caption{\label{fig:pt_tops13}%
    Transverse momentum distribution of the top quark (above) and
    anti-top quark (below) at 13 TeV.}
\end{figure}

We start comparing the transverse momentum distribution of the
reconstructed top and anti-top quarks, which is shown in
Figures~\ref{fig:pt_tops8} and~\ref{fig:pt_tops13}, for a center of mass energy of 8 TeV and 13 TeV respectively.

For either the signal or background process, the shapes of the top and
anti-top quark $p_T$ distributions are very similar, as expected, and
therefore the same is true for the signal-to-background ratio. In both cases,
it reaches a maximum between $50$ and $100~\GeV$ and than decreases
slightly in the high transverse momentum tail. 
Increasing the center-of-mass energy from 8 to 13 TeV does not 
lead to significant changes. Since this is true also for the other distributions that we 
studied, in the following we will only report and comment on the 13 TeV scenario.

It is worth noting, by looking at Figure~\ref{fig:pt_tops13},
that the uncertainty due to the PDF variation is larger in \ttH, due
to the dominant gluon-channel production, and is increasing for larger
transverse momenta. At $p_{T}\approx400~\GeV$ the PDF uncertainty for
\ttH{} is around 20\%, whereas it stays below 15\% for \ttyy.

\begin{figure}[t!]
  \centering
  \includegraphics[width=0.49\textwidth]{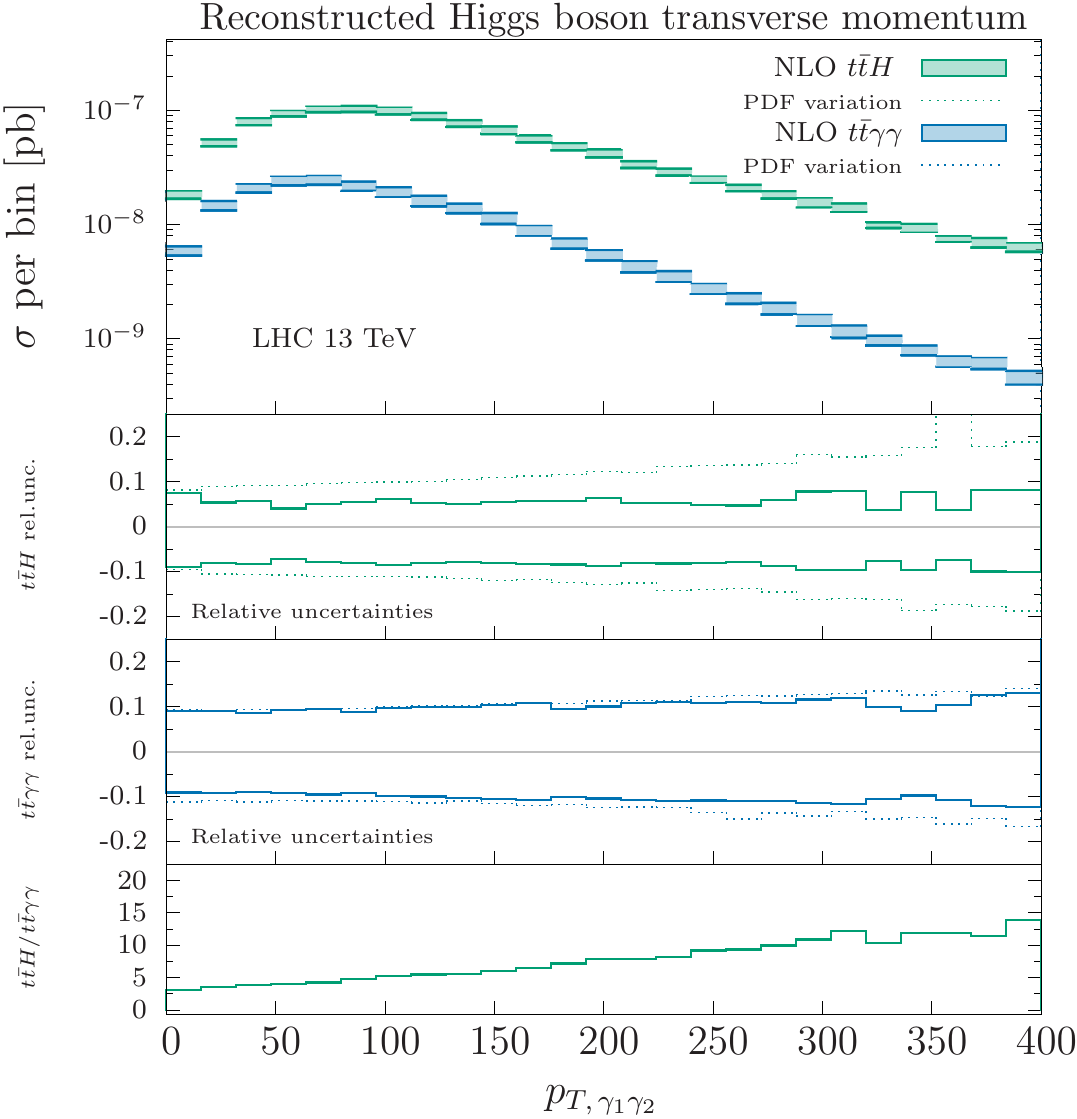}
  \includegraphics[width=0.49\textwidth]{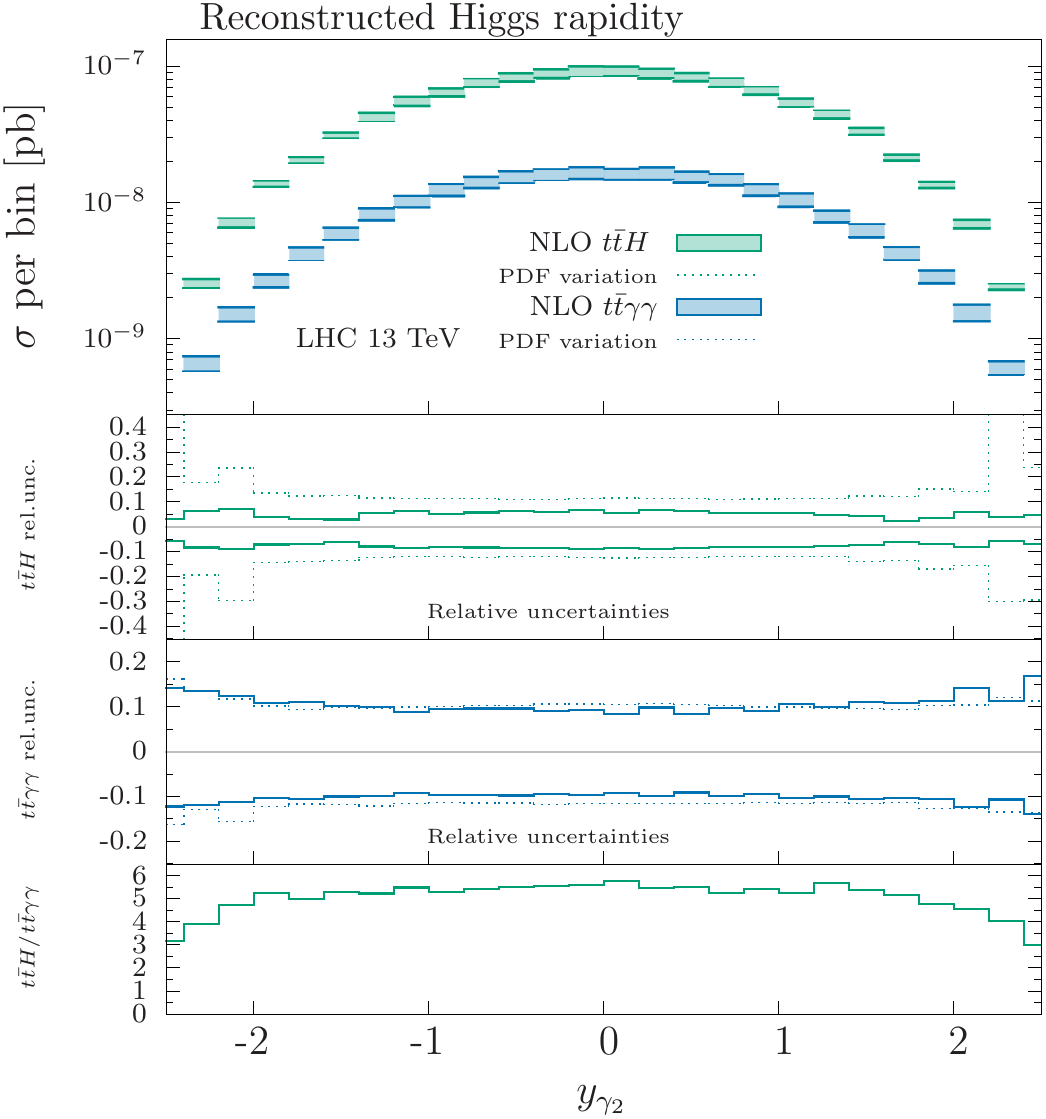}
  \caption{\label{fig:pt_mass_higgs}%
    Transverse momentum (top) and rapidity (bottom) distributions of
    the reconstructed Higgs boson.}
\end{figure}

Figure~\ref{fig:pt_mass_higgs} shows the transverse momentum and the
rapidity of the photon pair, which for the signal process corresponds
to the one of the reconstructed Higgs boson. Since in \ttyy{} the
photon pair does not originate from a massive particle decay, its
transverse momentum is softer and the spectrum falls off faster for
large $p_T$. The signal-to-background rapidity curve shows that
photons coming from the decay of the Higgs boson are generally
produced more centrally.  This is not surprising given that such a
decay does not feature a collinear enhancement, contrary to the case
of \ttyy.

\begin{figure}[t!]
  \centering
  \includegraphics[width=0.49\textwidth]{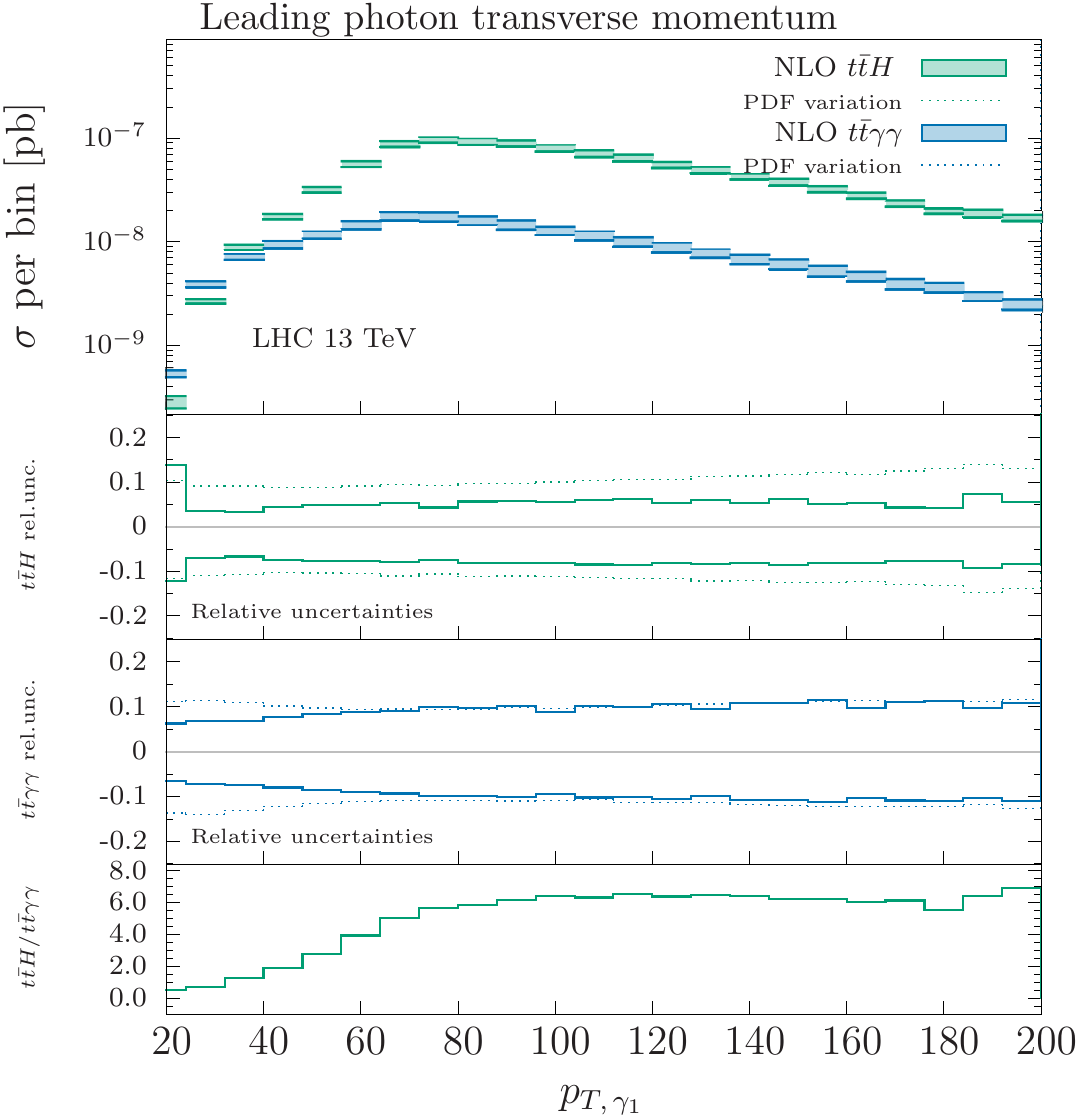}
  \hfill
  \includegraphics[width=0.49\textwidth]{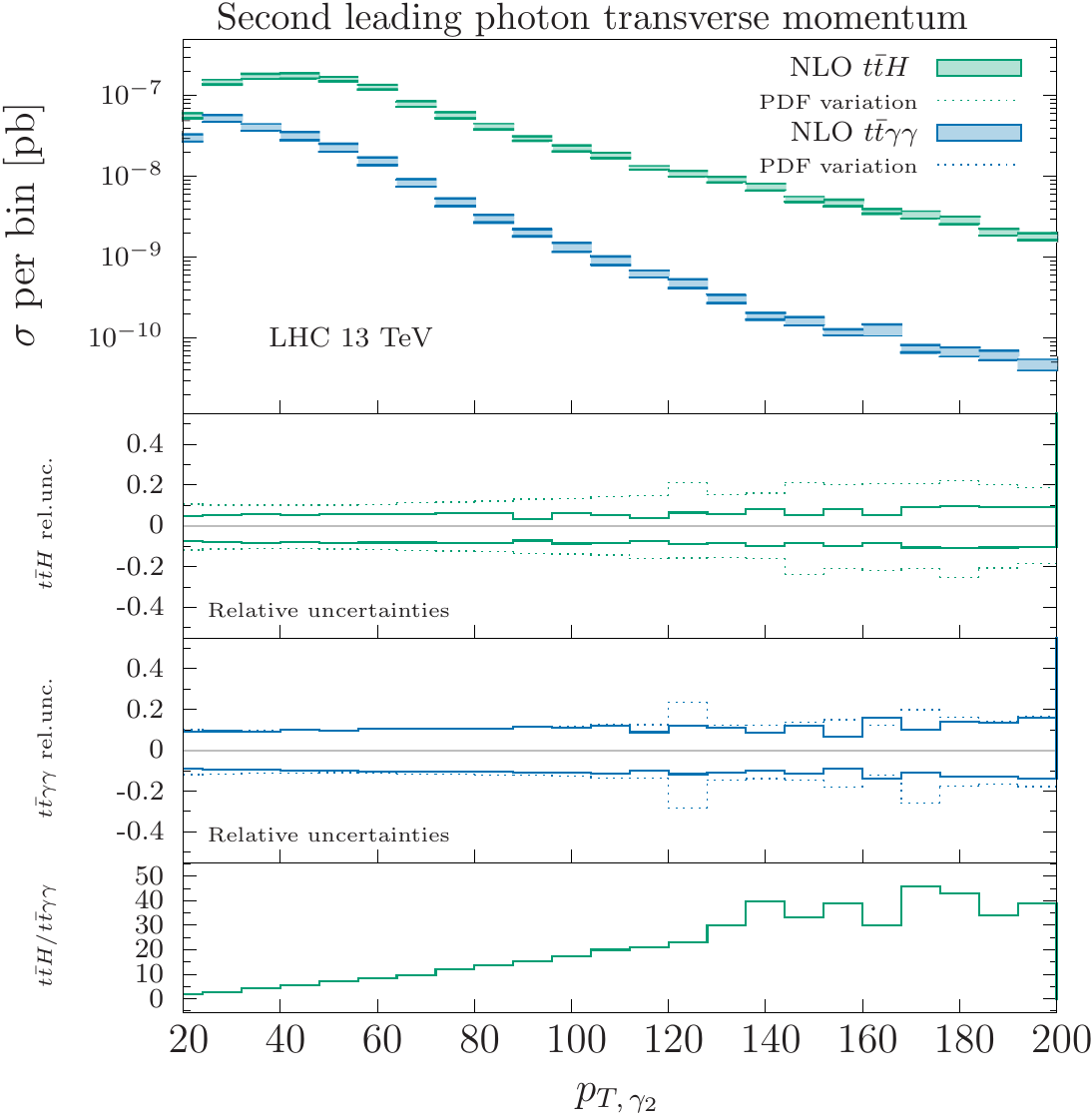}
  \caption{\label{fig:pt_gammas}%
    Transverse momentum distribution of the leading (top) and second
    leading photon (bottom).}
\end{figure}

The transverse momentum distribution for the single photons (ordered
according to their $p_T$) is shown in Figure~\ref{fig:pt_gammas}. The
shoulder in the \ttH{} signal distribution stemming from the presence
of the Higgs boson resonance is also visible in the background shape,
albeit less pronounced, given the invariant mass cut on the photon
pair. As expected, the shoulder is shifted towards lower transverse
momenta for the case of \ttyy, because of the initial state collinear
enhancement.

\begin{figure}[t!]
  \centering
  \includegraphics[width=0.49\textwidth]{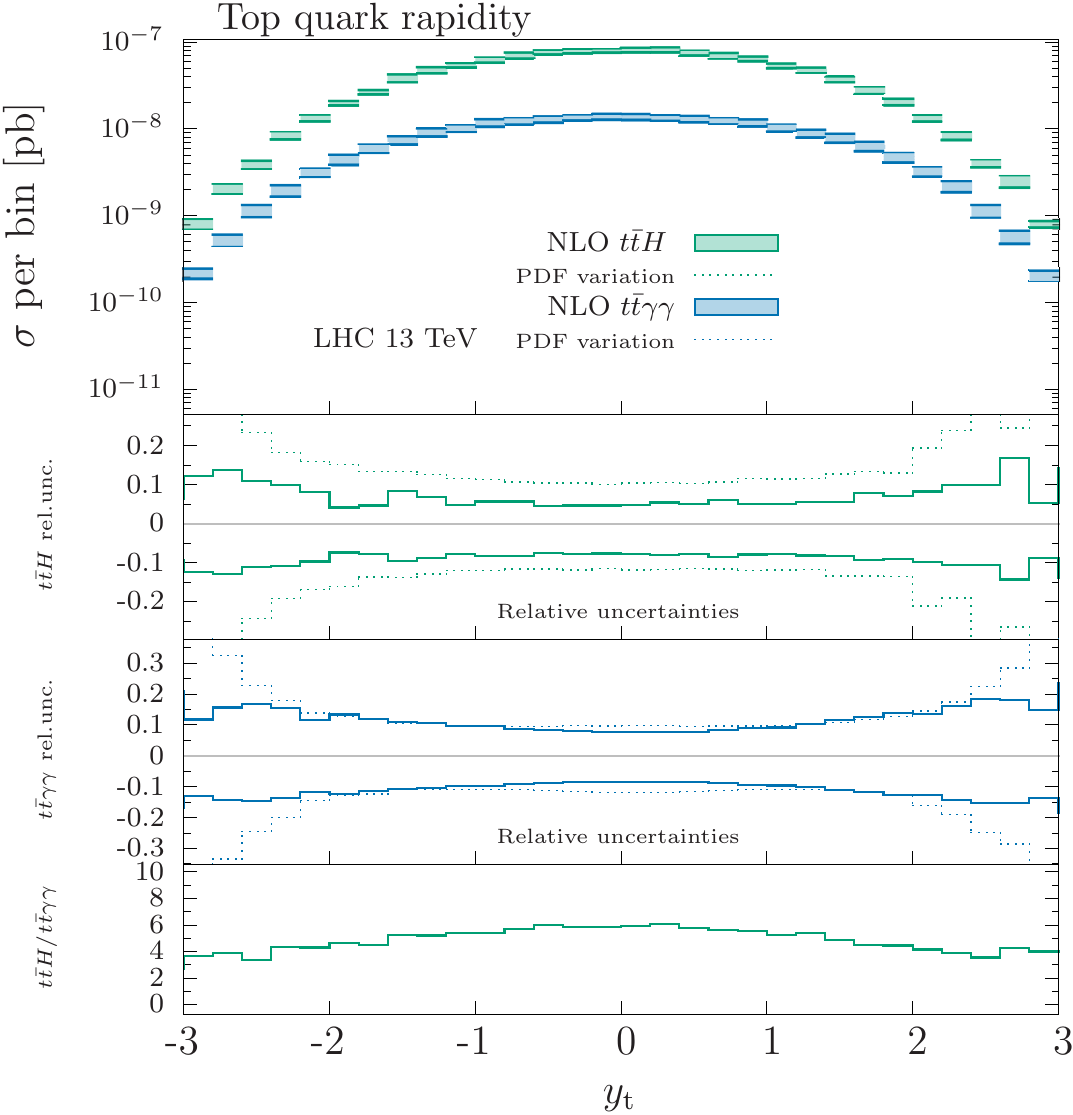}
  \hfill
  \includegraphics[width=0.49\textwidth]{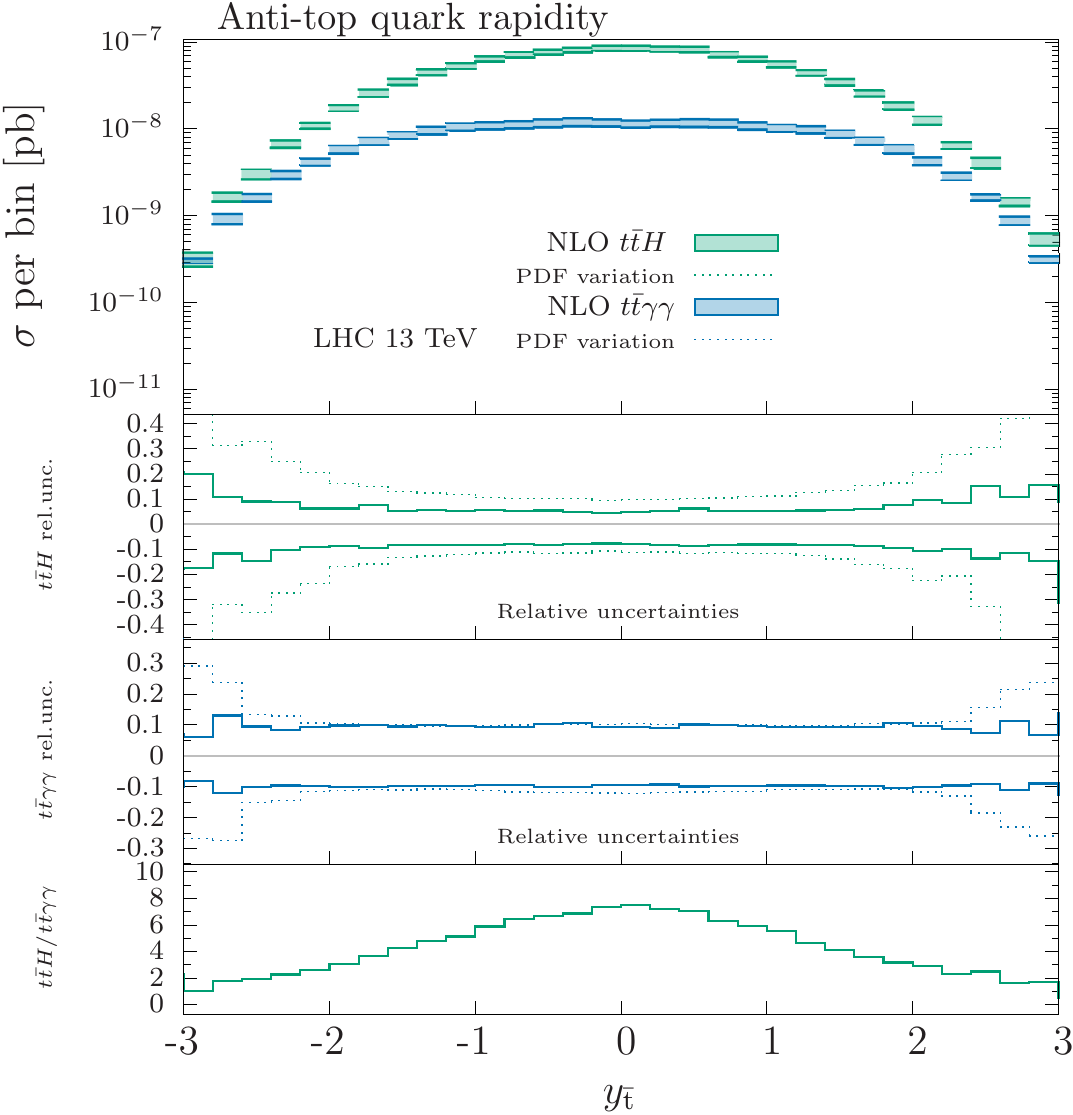}
  \caption{\label{fig:y_tops_SvsB}%
    Rapidity distribution of the top quark (top) and anti-top quark
    (bottom).}
\end{figure}

It is particularly interesting to compare the rapidities of the top
and anti-top quarks (Figure~\ref{fig:y_tops_SvsB}) and of their decay
products.  This highlights the well known difference between the
broadness of the respective top and anti-top quark rapidity
distribution and it can be used to improve on background
discrimination. This difference is known as the \emph{charge
  asymmetry} and is usually quantified by the following observable:
\begin{equation}
\label{eq:tt_asymmetry}
\Ac=\frac{\sigma\left(\Delta|y|>0\right)-\sigma\left(\Delta|y|<0\right)}
         {\sigma\left(\Delta|y|>0\right)+\sigma\left(\Delta|y|<0\right)}\,,
\end{equation}
where $\Delta|y|=|y_{\mathrm{t}}|-|y_{\bar{\mathrm{t}}}|$. This
observable has been measured at the LHC by both the ATLAS~\cite{ATLAS:2012an,Aad:2013cea} and the
CMS~\cite{Chatrchyan:2011hk,Khachatryan:2015oga,CMS:2014sfa} collaboration in the context of top-quark studies.

\begin{figure}[t!]
  \centering
  \includegraphics[width=0.47\textwidth]{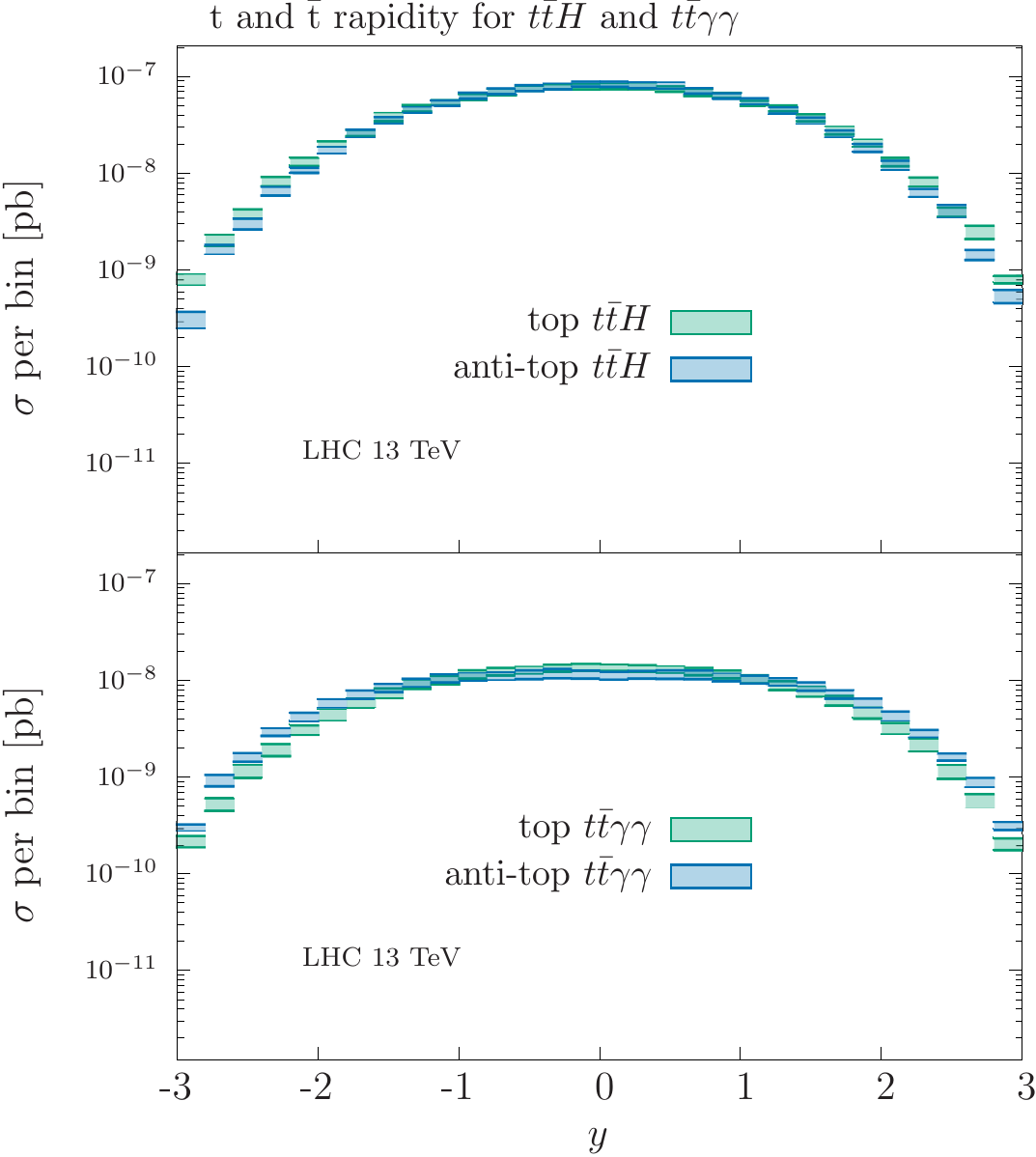}
  \hfill
  \includegraphics[width=0.47\textwidth]{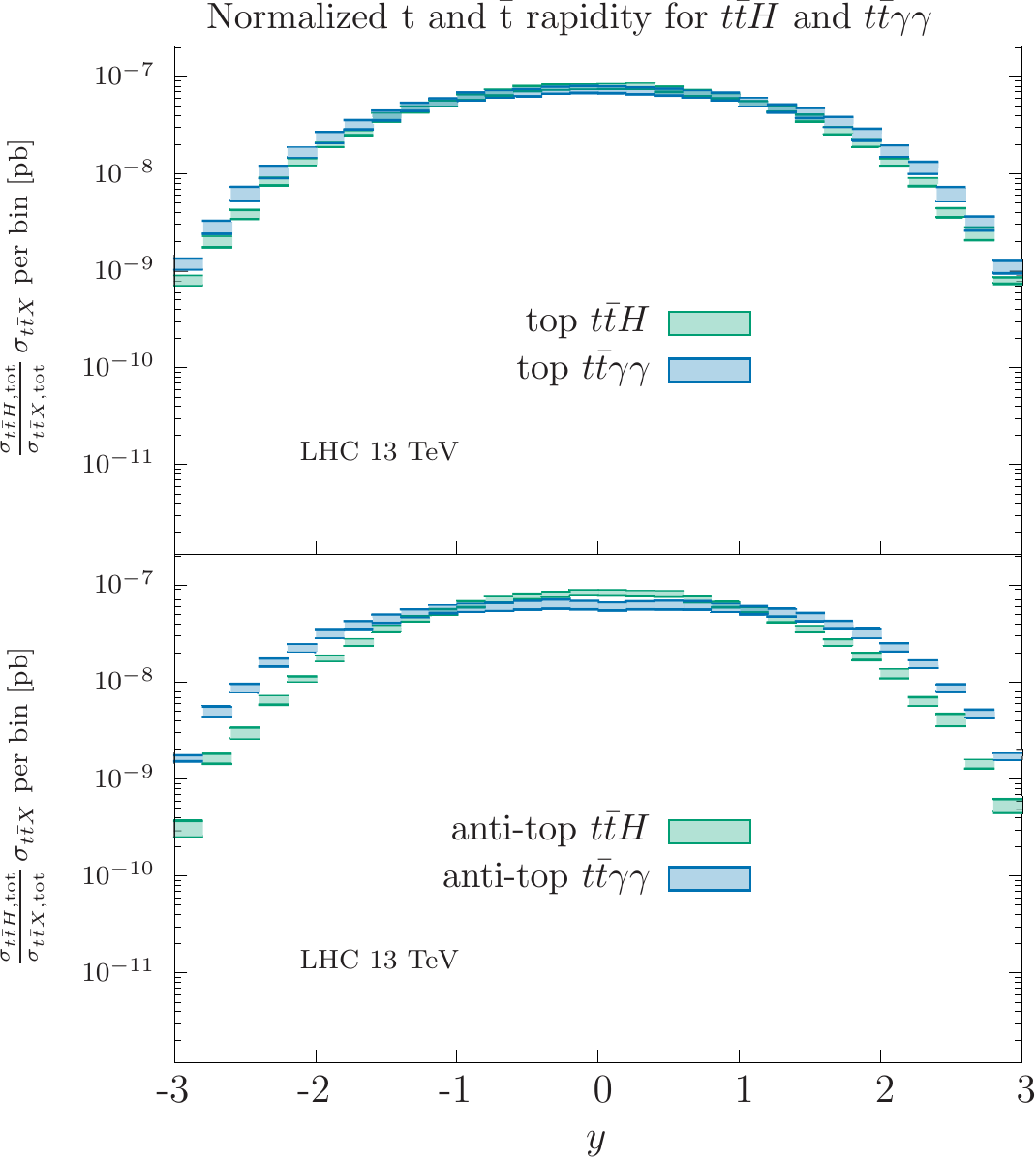}
  \caption{\label{fig:y_tops_ss_bb}%
    Upper plot: the rapidity distribution comparison between the top and
    anti-top quark for \ttH{} (above) and \ttyy{} (below). Lower plot:
    normalized top (above) and anti-top (below) rapidity distribution
    for signal and background.}
\end{figure}

For top-pair production at the LHC $\Ac$ is positive, i.e.~top quarks
are produced at larger rapidities compared to anti-top
quarks~\cite{Kuhn:1998kw}. It is however known that the presence of a
photon reverses the sign of $\Ac$ already at
tree-level~\cite{Aguilar-Saavedra:2014vta}. This change in the
rapidity distributions of the top and anti-top quarks, due to
additional photon radiation, can clearly be seen in the plots of
Figure~\ref{fig:y_tops_ss_bb}, which compare $y_\mathrm{t}$ and
$y_{\mathrm{\bar{t}}}$ individually for \ttH{} and \ttyy. In the
signal process the additional presence of a Higgs boson in the final
state does not change the qualitative result, as compared to simple
$t\bar{t}$-production. This can be seen in the upper portion of the top plot in
Figure~\ref{fig:y_tops_ss_bb}, which shows that top quarks are
produced at slightly higher rapidities, as compared to anti-tops, which
are more central, leading to a positive $\Ac$. In the lower left plot,
instead, the effect is reversed. The presence of the additional
photons causes the top quarks to be more central compared to the
anti-tops.

This effect is even more visible when comparing directly the
distributions for \ttH{} with the ones for \ttyy{}. To better
appreciate the change in the shape, which is only marginally visible
in Figure~\ref{fig:y_tops_SvsB}, we plot the same distribution
normalized to the inclusive \ttH{} cross section on the lower part of
Figure~\ref{fig:y_tops_ss_bb}. From the upper plot it becomes clear
that the top quark rapidities have very similar shape in both the
signal and the background process, although in the latter the tops are
produced at slightly higher rapidities. This means that despite the
top being produced at higher rapidities as compared to the anti-top in
\ttH, overall, they are still slightly more central than in \ttyy. The
opposite is true for the anti-top quark rapidity, and, therefore, the
difference is even more visible in the lower plot of
Figure~\ref{fig:y_tops_ss_bb}.

\begin{figure}[t!]
  \centering
  \includegraphics[width=0.49\textwidth]{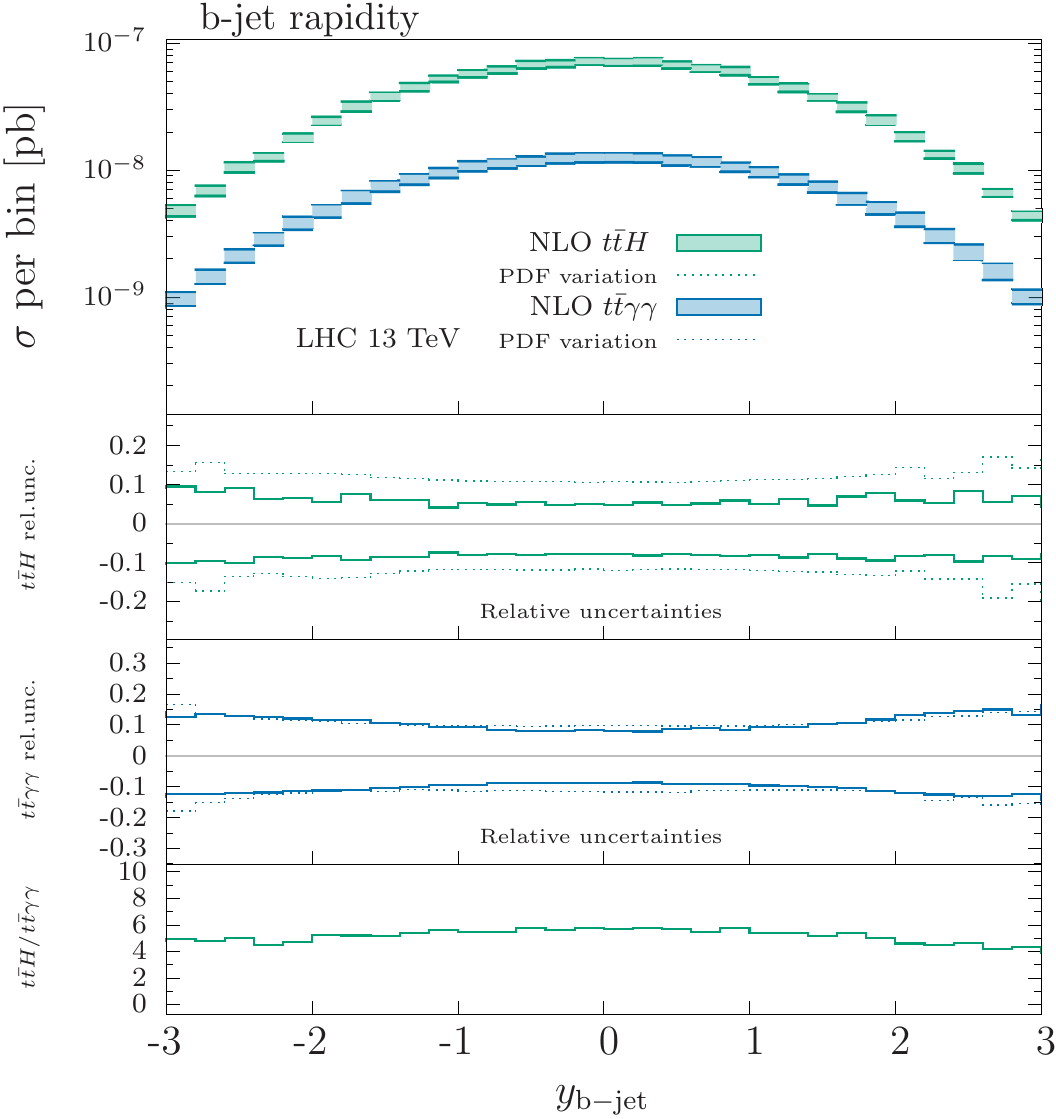}
  \hfill
  \includegraphics[width=0.49\textwidth]{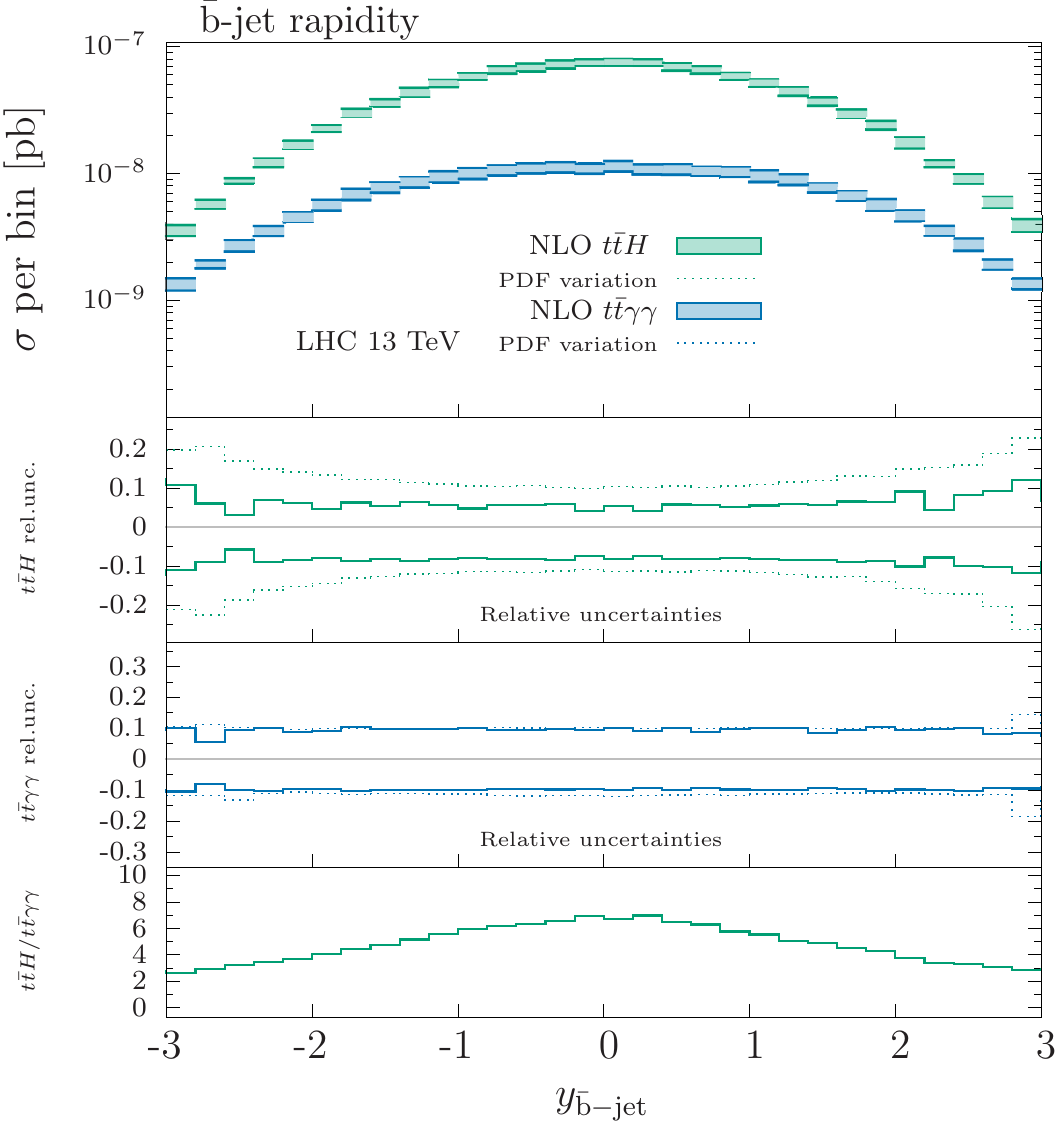}
  \caption{\label{fig:y_bottoms_SvsB}%
    Rapidity distribution distribution of the b-jet (top) and
    anti-b jet (bottom) produced in the decay of the top and
    anti-top quark respectively.}
\end{figure}

\begin{figure}[t!]
  \centering
  \includegraphics[width=0.49\textwidth]{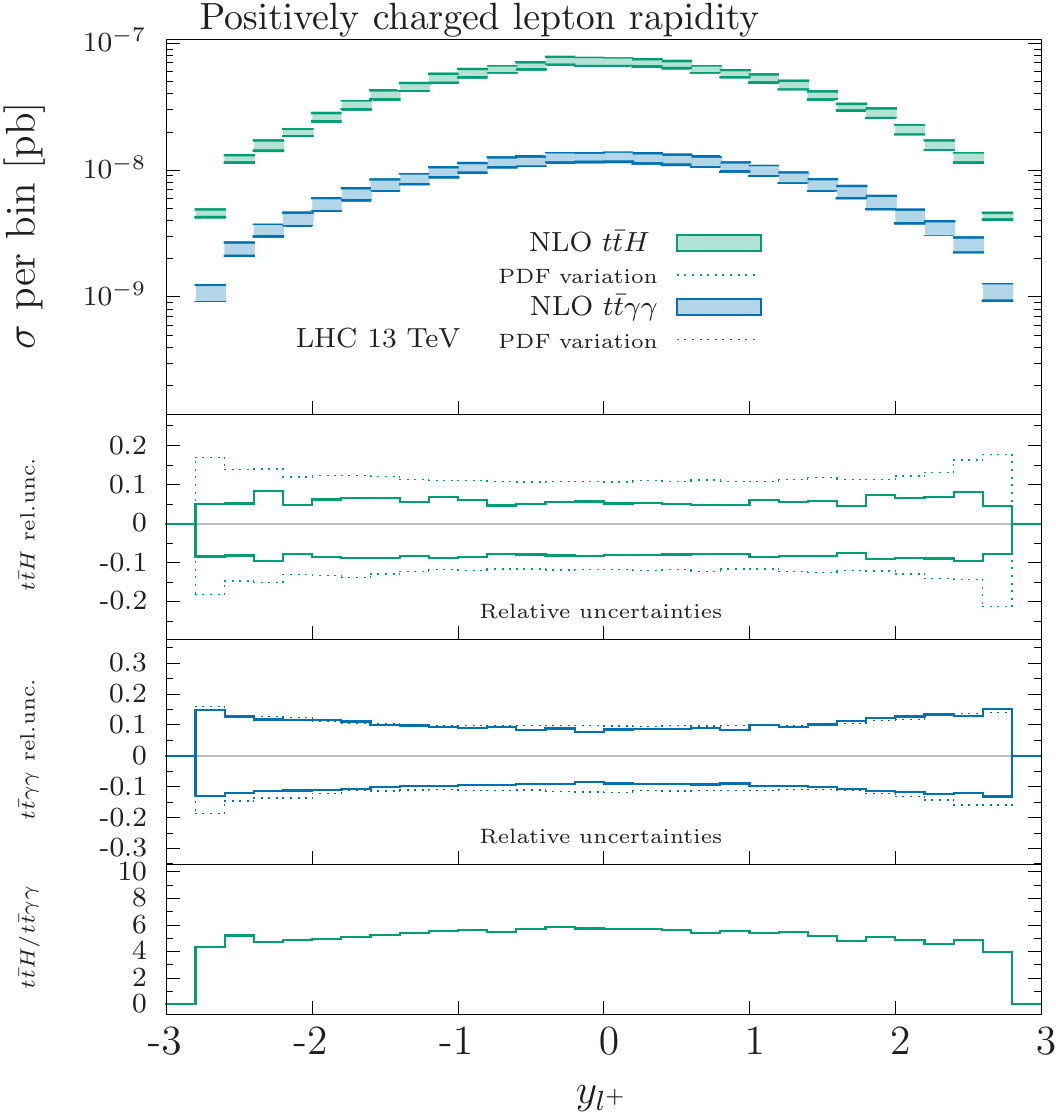}
  \hfill
  \includegraphics[width=0.49\textwidth]{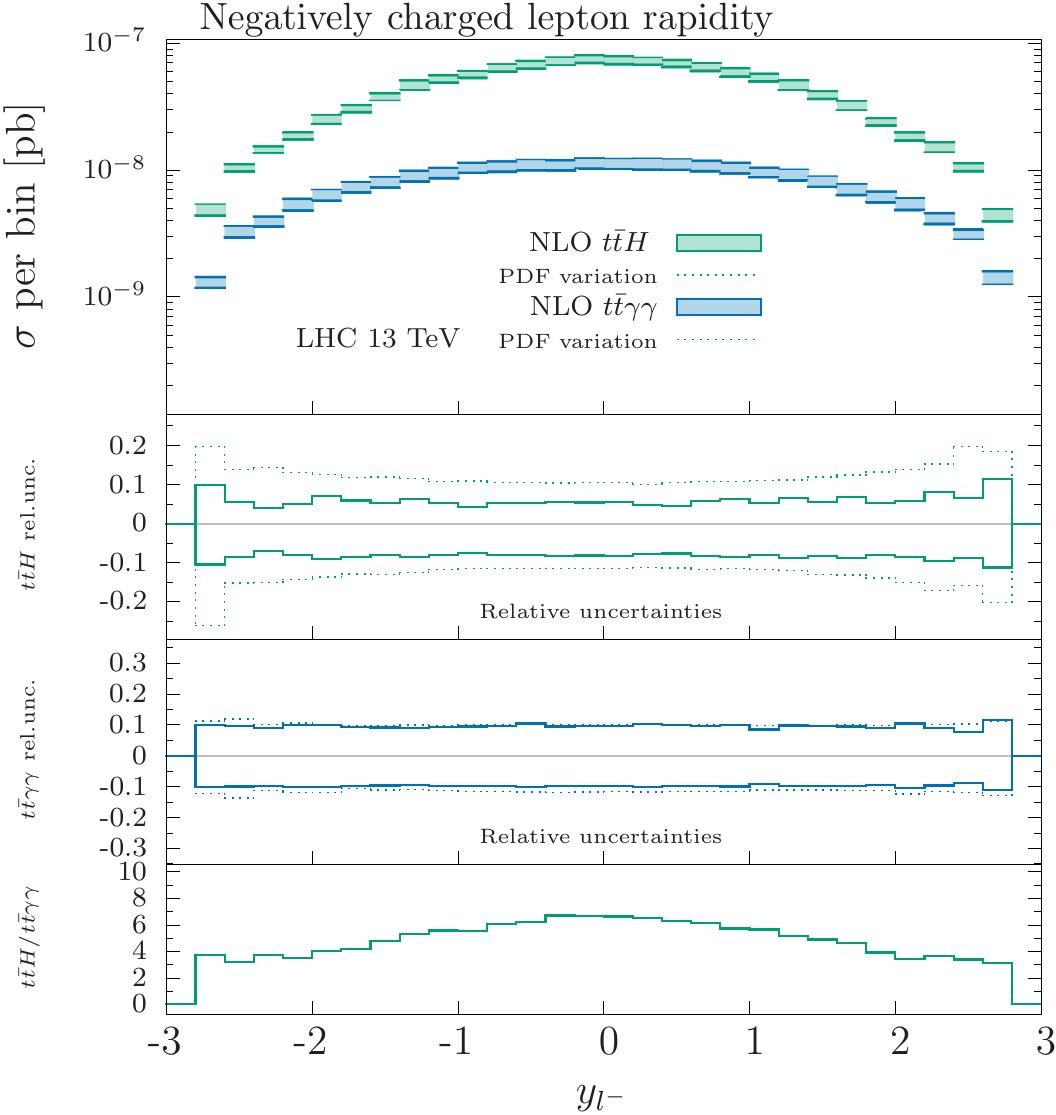}
  \caption{\label{fig:y_leptons_SvsB}%
    Rapidity distribution of the lepton (top) and
    anti-lepton (bottom) produced in the decay of the top and
    anti-top quark respectively.}
\end{figure}

Analogous conclusions can be derived by looking at the rapidities of
the top decay products. They are shown in
Figure~\ref{fig:y_bottoms_SvsB} and~\ref{fig:y_leptons_SvsB}, where
the rapidities of the $b$- and $\bar{b}$-jets and the charged
leptons are shown respectively.

\begin{figure}[t!]
  \centering
  \includegraphics[width=0.49\textwidth]{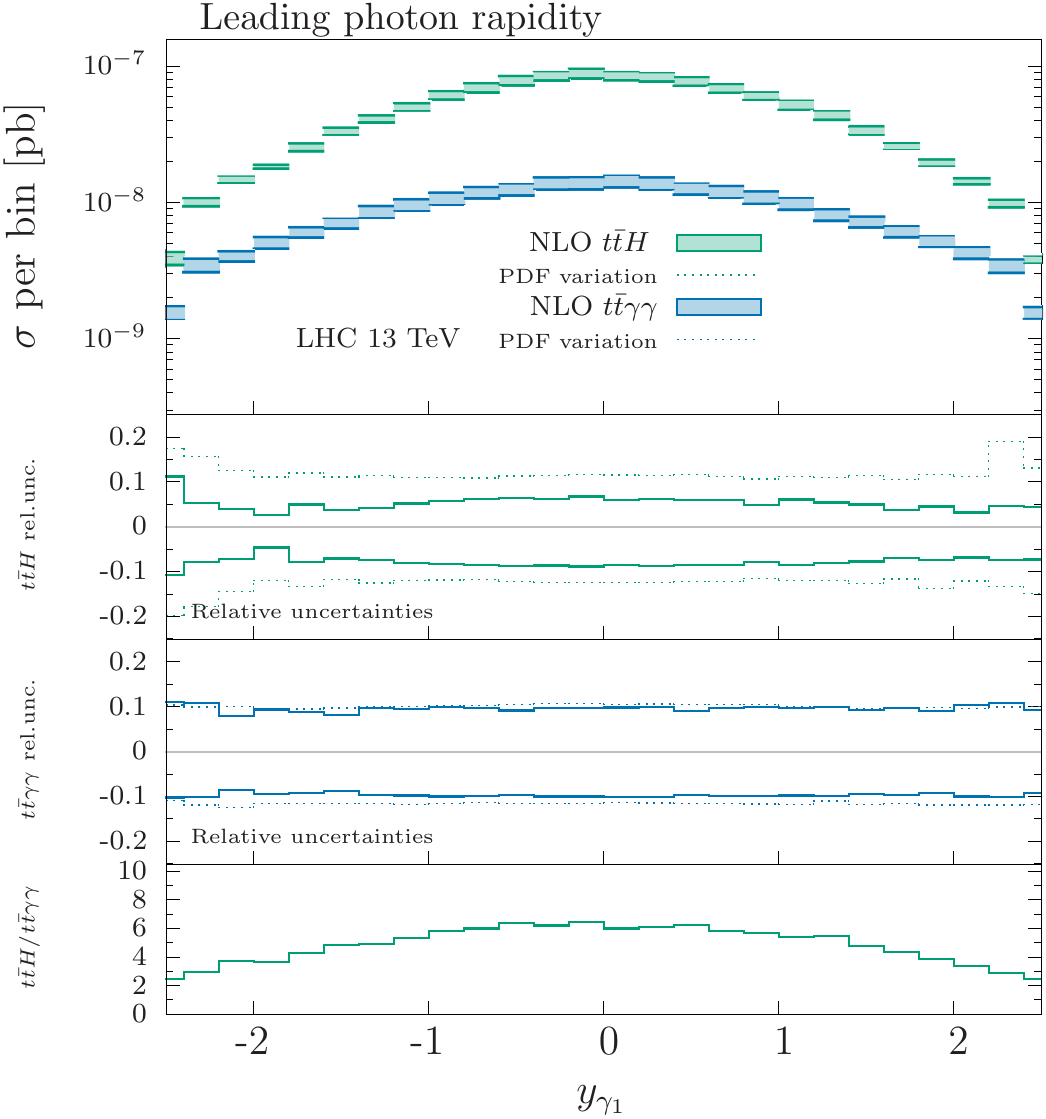}
  \hfill
  \includegraphics[width=0.49\textwidth]{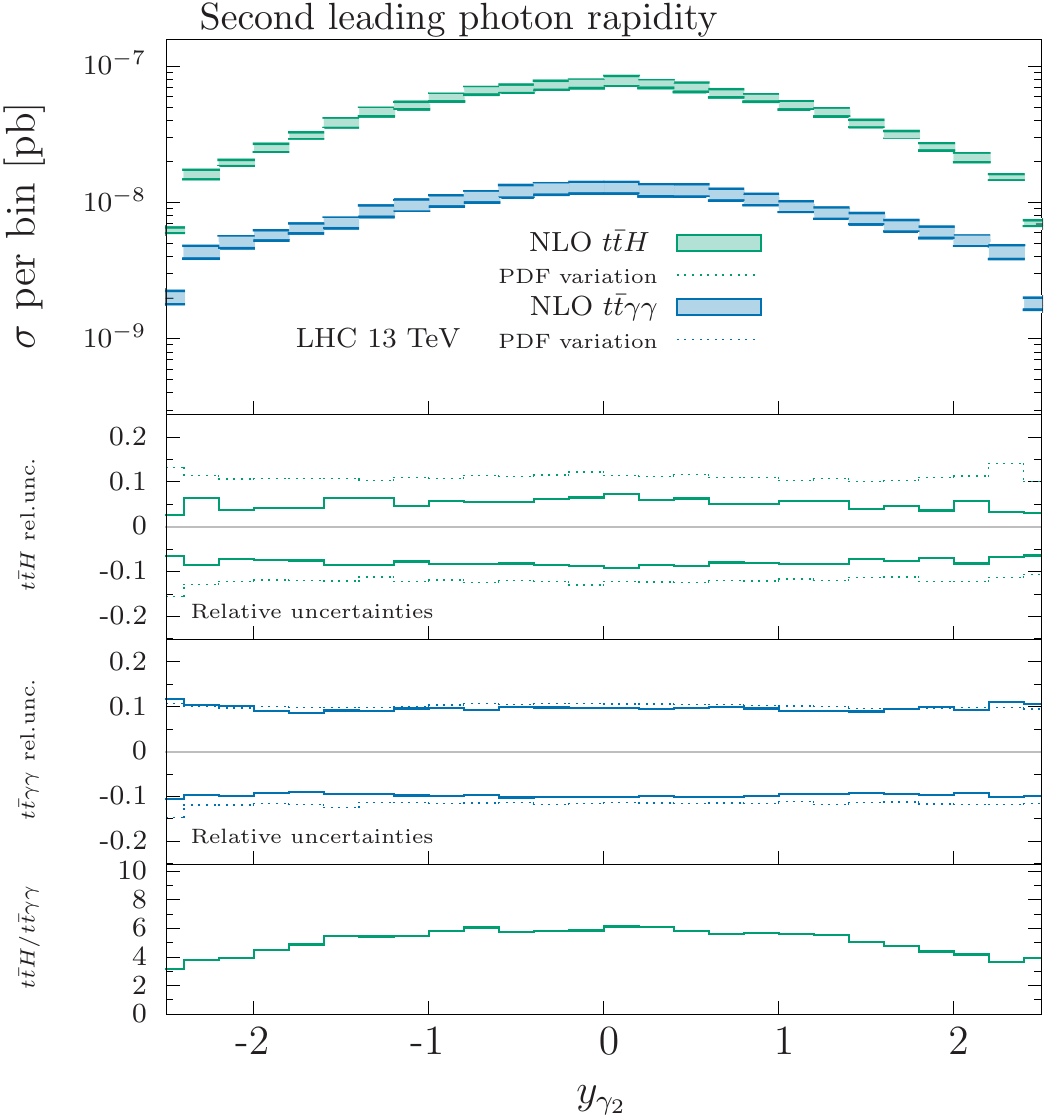}
  \caption{\label{fig:y_gammas}%
    Rapidity distribution of the leading (top) and
    second leading photon (bottom).}
\end{figure}

In Figure~\ref{fig:y_gammas} we compare the rapidities of the leading
and second leading photon. Not surprisingly, the photon coming form the
Higgs boson decay are produced more centrally as compared to the ones
radiated from the partons.

\subsection{Spin polarisation observables}
\label{Subsec:spinobservables}
In this section we focus on observables that allow for investigating
polarisation effects of the top and anti-top quarks as well as in their decay
products. This can be done by studying angular variables which involve
the decay products of the top and anti-top quarks; both top quarks are considered to decay semileptonically (same as in the previous section). We stress that a similar analysis was already
performed at LO in Ref.~\cite{Biswas:2014hwa}.

\begin{figure}[t!]
  \centering
  \includegraphics[width=0.49\textwidth]{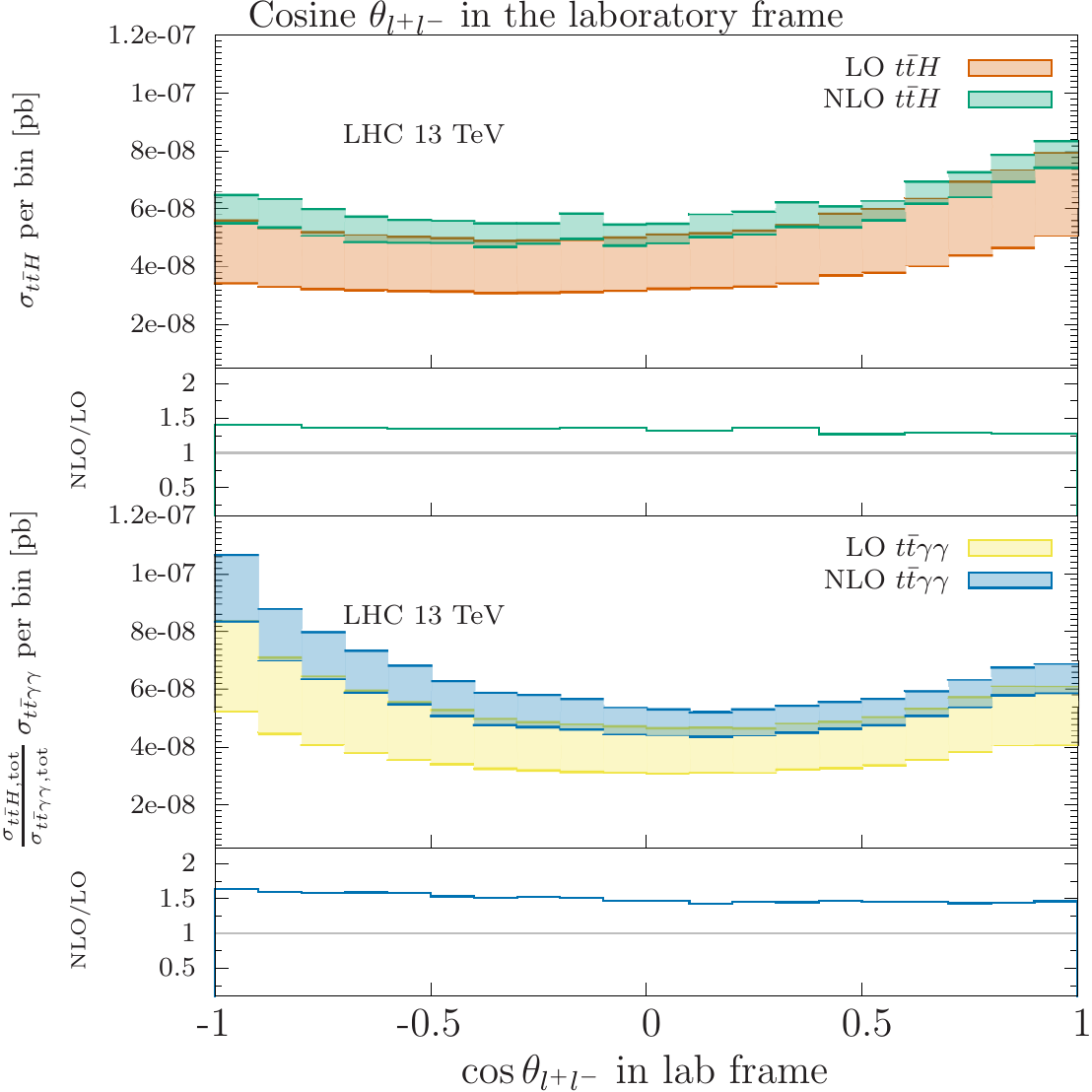}
  \hfill
  \includegraphics[width=0.49\textwidth]{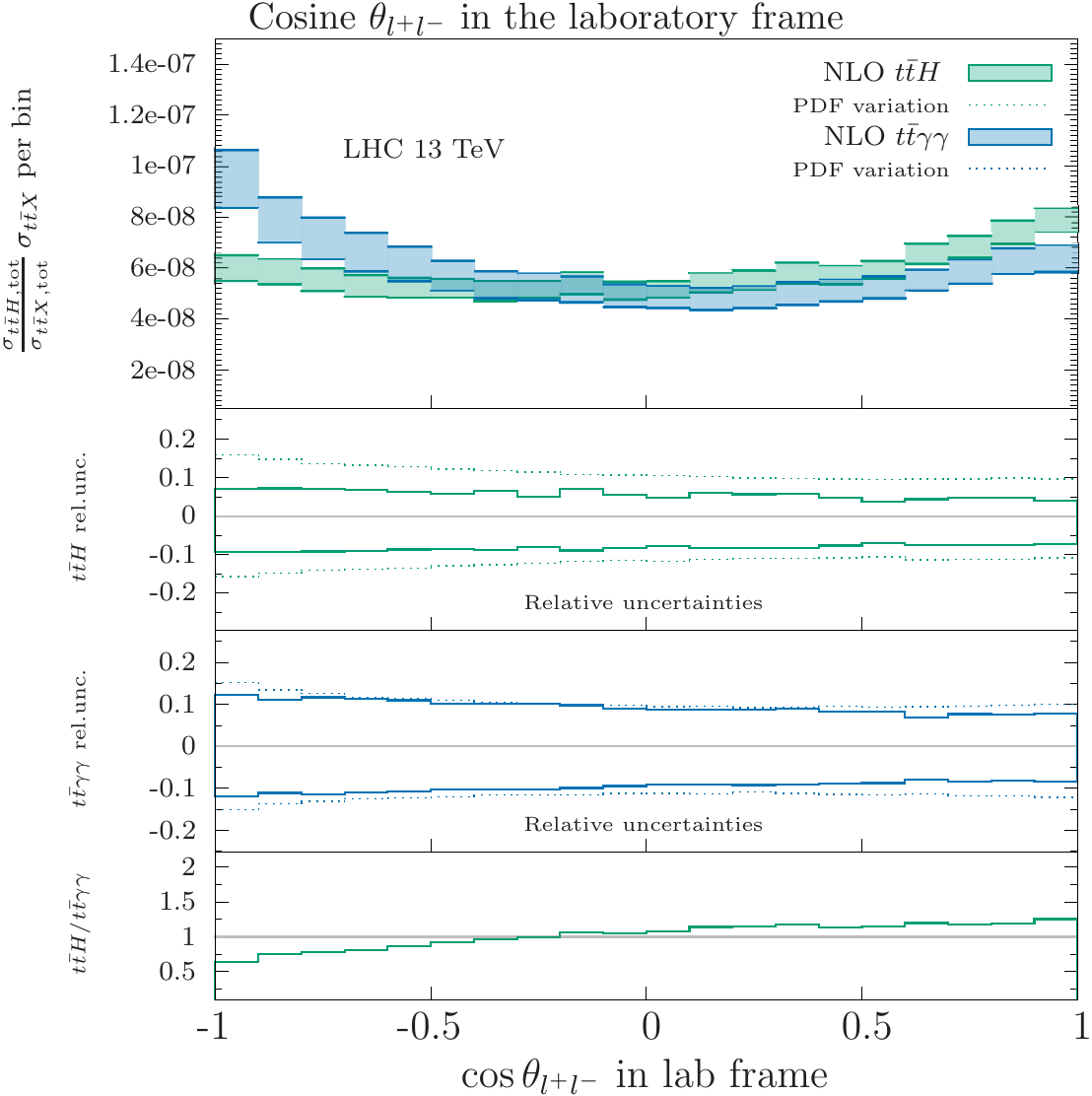}
  \caption{\label{fig:costhlplm_lab}%
    $\cos\theta_{ll}$ distribution for the signal (\ttH) and
    background (\ttyy) processes in the laboratory frame. The exact
    definition of the angle $\theta$ is given in the text. The \ttyy{}
    prediction is normalized to the \ttH{} inclusive cross-section. In the upper plot, 
    we compare LO with NLO predictions and show their
    K-factor separately for \ttH{} and \ttyy{} in bottom insets. In the lower plot,
    we show NLO relative uncertainties with the signal-to-background ratio as the last bottom inset.}
\end{figure}

Typically, for hadronic $t\bar{t}$-production, very specific kinematic frames are
defined~\cite{Bernreuther:2001rq,Bernreuther:2004jv,Bernreuther:2010ny}.
In the following we will consider the three-dimensional opening angle
$\theta_{ll}$ between the leptonic decay products of the top ($l^+$)
and anti-top quarks ($l^-$), defined in three different frames. The
most straightforward possibility is to define $\theta_{ll}$ in the
laboratory frame (referred to as lab-frame in the following). The
results for this case are shown in Figure~\ref{fig:costhlplm_lab}.
The definition of two other frames, introduced for the first
time in~\cite{Bernreuther:2004jv}, became customary in polarisation
studies, since they capture particularly well spin correlation
effects.

For these particular frames, we define $\theta_{ll}$ to be the angle
between the direction of flight of $l^+$, measured in frame where
the \emph{top quark} is at rest, and the direction of flight of $l^-$,
measured in the frame where the \emph{anti-top quark} is at rest.
Since two rest frames are involved in this definition, a common
initial frame needs to be specified, from which the (rotation-free)
Lorentz boost can be applied in order to transform the system to the
$t$ and $\bar{t}$ rest frames. We choose two possible starting points,
which we label as \textit{frame-1} and \textit{frame-2},
defined as follows:

\begin{itemize}
  \item{\textit{frame-1}: the Lorentz boosts to bring $t$ and $\bar{t}$
    separately at rest are defined with respect to the $t\bar{t}$-pair
    center-of-mass frame,}
  \item{\textit{frame-2}: the Lorentz boosts to bring $t$ and $\bar{t}$
    separately at rest are defined with respect to the lab-frame.}
\end{itemize}

These two frames are designed to be maximally sensitive to the
different polarisation structures of the top-pair in the final
state. Furthermore, as already demonstrated at
LO~\cite{Biswas:2014hwa}, considering the spin information in the
decay of the top and anti-top quark is crucial to disentangle the two
different final states, which otherwise look identical, being
characterized by a completely flat distribution in both cases.

\begin{figure}[t!]
  \centering
  \includegraphics[width=0.49\textwidth]{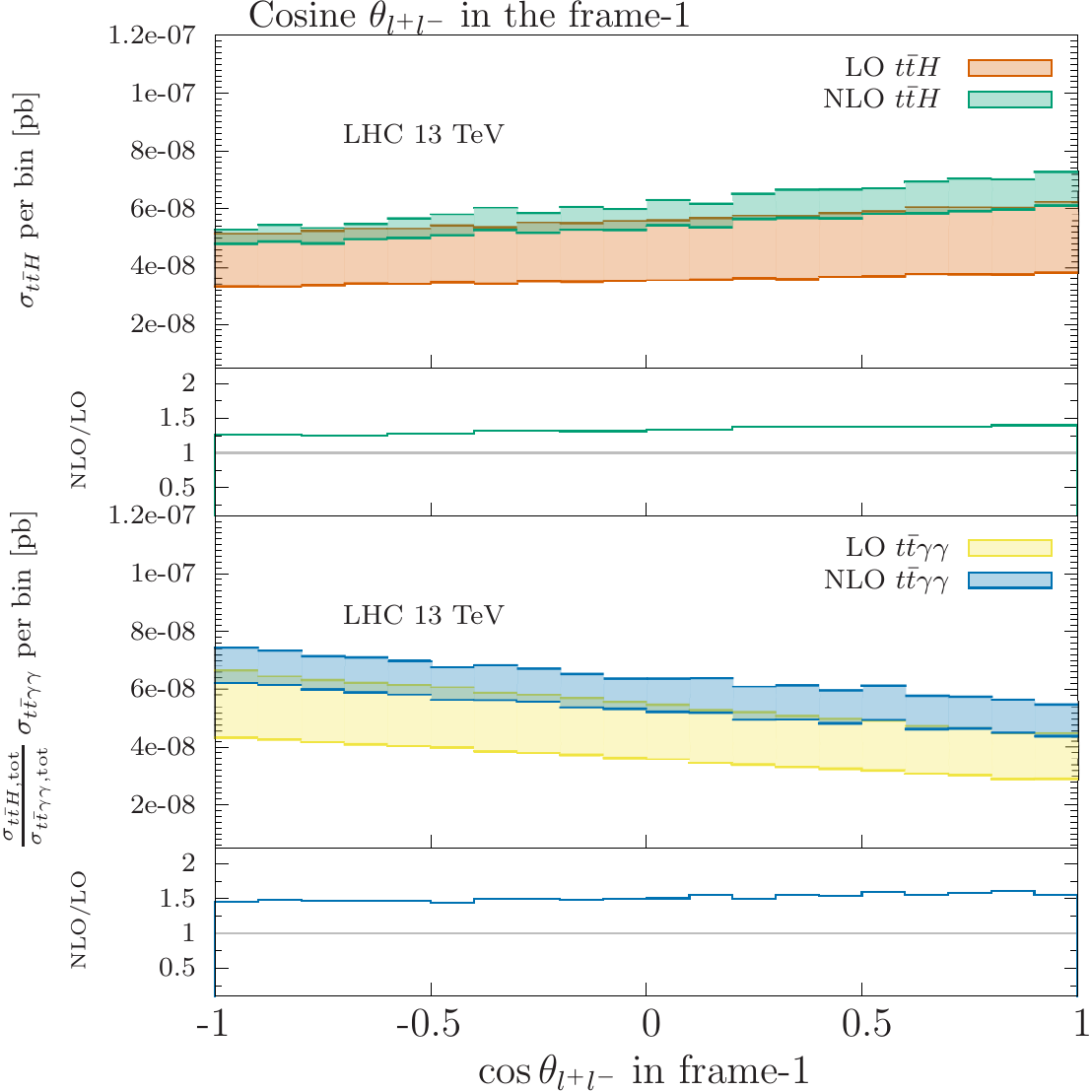}
  \hfill
  \includegraphics[width=0.49\textwidth]{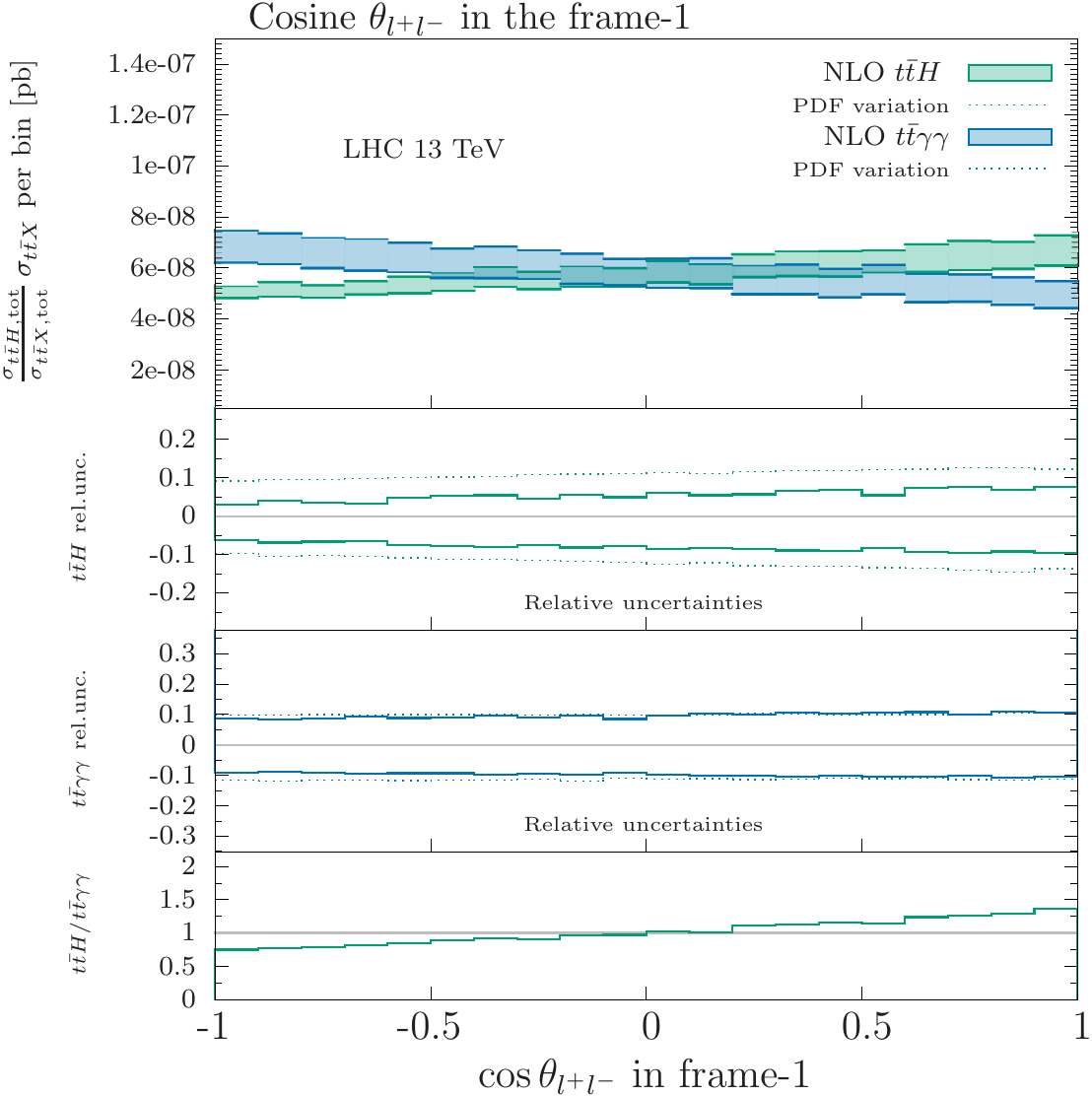}
  \caption{\label{fig:costhlplm_frame1}%
    Same as Figure~\ref{fig:costhlplm_lab}, but for reference frame-1.
    }
\end{figure}

\begin{figure}[t!]
  \centering
  \includegraphics[width=0.49\textwidth]{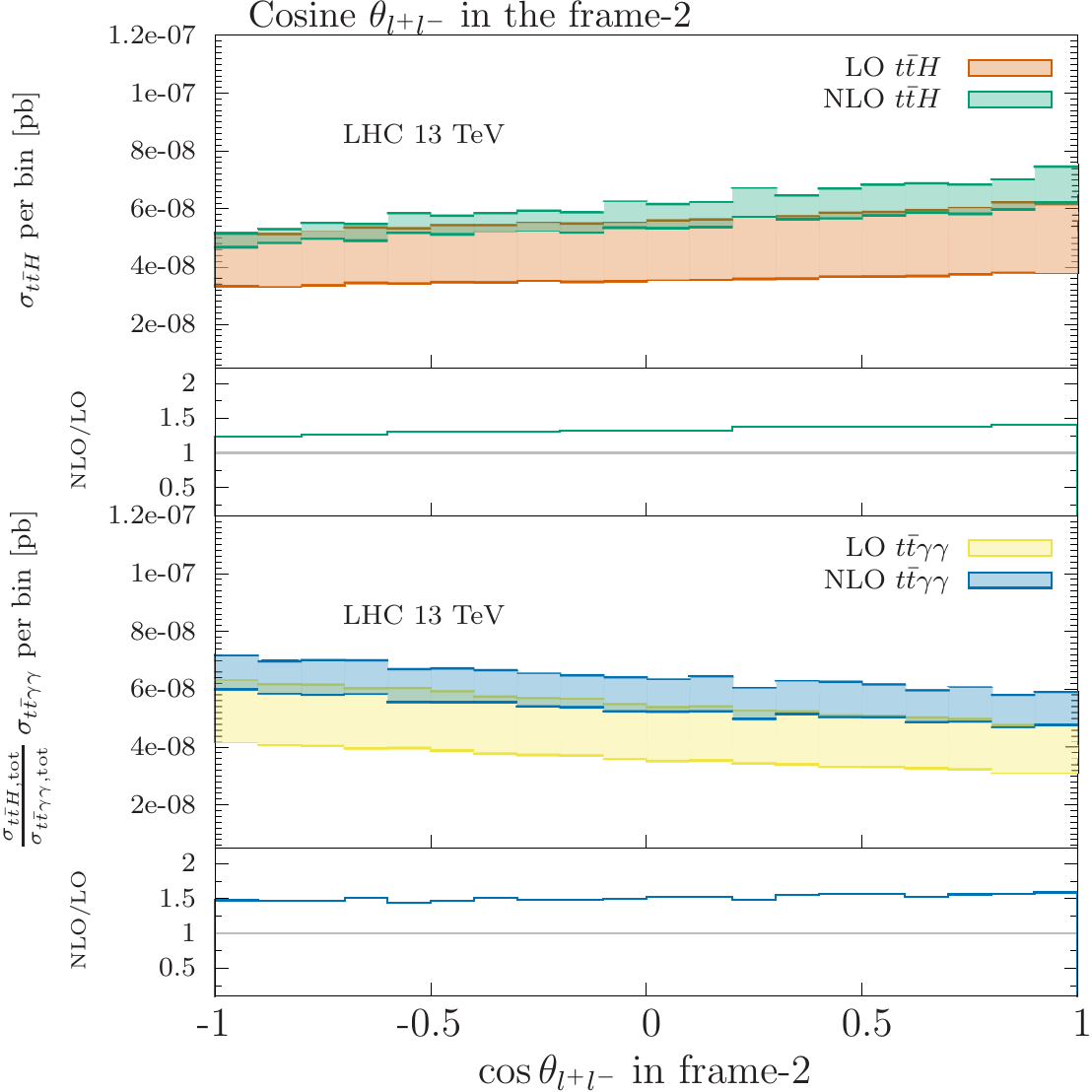}
  \hfill
  \includegraphics[width=0.49\textwidth]{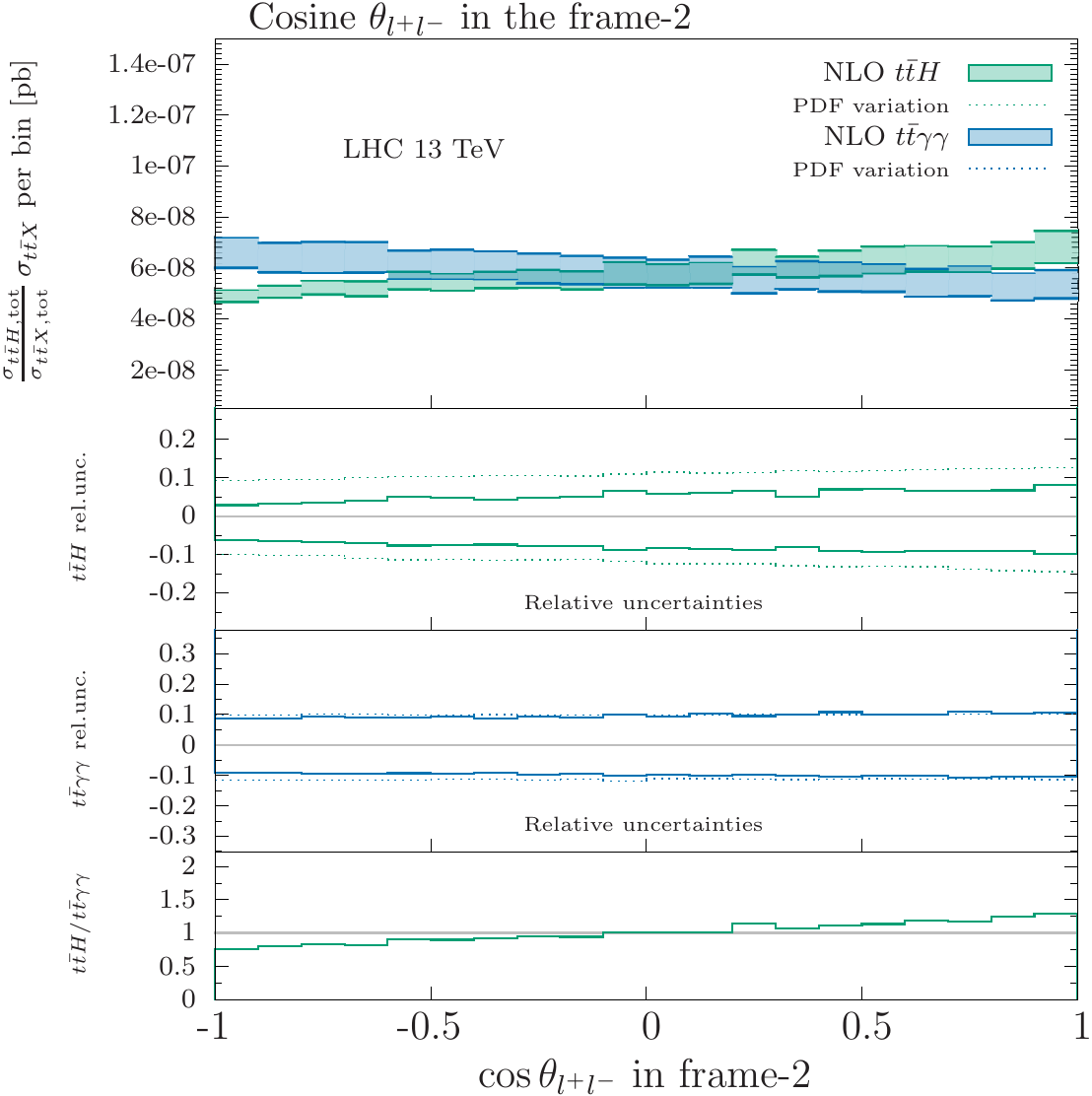}
  \caption{\label{fig:costhlplm_frame2}%
    Same as Figure~\ref{fig:costhlplm_lab}, but for reference frame-2.
    }
\end{figure}

In Figure~\ref{fig:costhlplm_lab} we show the behaviour of
$\cos\theta_{ll}$ in the lab-frame. To highlight shape differences,
the background predictions have been normalized to the inclusive
\ttH{} cross section.  The upper portions of 
Figures~\ref{fig:costhlplm_lab}-\ref{fig:costhlplm_frame2} presents
the comparison of the LO and NLO predictions for \ttH{} and \ttyy{}
separately and shows their respective differential K-factors. The histograms in the lower portion of the 
Figures compare results for the signal and background
processes, with their ratio and respective relative uncertainty in the
bottom insets.

This is to be compared with the plots in
Figures~\ref{fig:costhlplm_frame1}-\ref{fig:costhlplm_frame2}, where
the same observable is shown in frame-1 and frame-2. In the two latter
frames a difference in the sign of the slope emerges, while in the lab
frame, despite a clear difference in the slopes, the curves have an
analogous trend. By comparing the two ratio plots at the bottom of the
right plots in Figures~\ref{fig:costhlplm_frame1}
and~\ref{fig:costhlplm_frame2}, we conclude that the
frame-1 offers the best signal-to-background ratio. It is also worth
stressing that, while in the lab frame the K-factors tend to decrease
slightly when $\cos\theta_{ll}\to 1$, in frame-1 and frame-2, the NLO
corrections feature an almost perfectly flat K-factor which agrees
with the inclusive cross section K-factor reported in
Table~\ref{table:xstable8}. A comparison of the LO and NLO
predictions reveals the anticipated reduction of the scale
uncertainties.

Let us finally remark that the results shown here for the background,
only consider photon radiation from the initial state and from the top
and anti-top quark, but not from their decay products. This is
opposite to the case of the signal, where photons always originate from
the decay of the Higgs boson. By considering more general cases and using
top tagging techniques without relying on MC truth is expected to
decrease the purity of the signal.  A more quantitative analysis of
these effects is beyond the scope of the present work.

\section{Conclusions}
\label{Sec:conclusions}
The event generator \amcnlomg{} and the one-loop amplitudes
provider \Gosam{} have been interfaced to provide the user with a
framework implenting the most advanced techniques for the evaluation
of cross sections and differential distributions at next-to-leading
order (NLO) accuracy.  

In this work, the integration of the two codes has been applied for
the first time to the NLO corrections to the production of a Higgs
Boson in association with a pair of top-antitop quarks, as well as to
the background process where two hard photons are produced directly. We
compared several key distributions to disentangle the two processes
and focused in particular on observables designed to study spin
correlation effects. We found that NLO corrections give a sizable
contribution, which however distorts the shape of the distributions
only very mildly. Moreover, we observed a clear reduction of
the theoretical uncertainties.

The high-level of flexibility and reliability of the joined
technologies of the two codes make of the combination of \amcnlomg{} and \Gosam{} an ideal
tool for the high-precision studies and the hunt for deviations from
known-physics signals which characterise the Run II programme at the
LHC.

\begin{acknowledgements}
We thank German Rodrigo and Jan Winter for valuable discussions. The
work of H.v.D., R.F., G.L.~and~P.M.~is supported by the Alexander von
Humboldt Foundation, in the framework of the Sofja Kovaleskaja Award
Projects ``Advanced Mathematical Methods for Particle Physics''
(H.v.D., G.L.~and~P.M.) and ``Event Simulation for the Large Hadron
Collider at High Precision'' (R.F.), endowed by the German Federal
Ministry of Education and Research. The work of G.O.~is supported in
part by the NSF under Grants PHY-1068550 and PHY-1417354. V.H.~is
supported by the Swiss National Fund for Science (SNSF) under grant
number P300P2\_161050.

This research made use of the CTP computational cluster of the New
York City College of Technology. We thank the CP3 IT team for their
constant support and the availability of the cluster hosted by the Universit\'e Catholique de Louvain.
\end{acknowledgements}

\appendix

\section{The \Gosam{} input card}

We report here a copy of the default \Gosam{} input card with a brief
explanation of the different options. For a more detailed overview we
refer to the \Gosam{} papers~\cite{Cullen:2011ac, Cullen:2014yla} and
the \Gosam{} manual which can be found online~\cite{gosamhome} and is
continuously updated.  The default
\texttt{gosam.rc} file is the following:

\begin{minipage}[t]{.45\textwidth}
\footnotesize{
\begin{Verbatim}[frame=single]
###############################################
# Copy this file to setup.in in order to set
# some common options for all examples.
###################
# physics options #
###################
# Model specs.:
model=smdiag_mad
model.options=GF: 0.0000116639, mZ: 91.188, 
mW: 80.419, Nf: 4

# Parameters set to zero algebraically:
zero=me,mmu,mU,mD,mC,mS,wB,wT

# Symmetries:
symmetries=family,generation

# Filter for scale-less loop integrals:
filter.nlo=lambda d: (not (d.isScaleless()))

###################
# program options #
###################
form.bin=tform
form.threads=4
form.tempdir=/tmp
fc.bin=gfortran -O2
###############################################
\end{Verbatim}
}
\end{minipage}
~\\

This input file needs to be modified if the computation is performed within the 5-flavour scheme. 
The b-quark mass can be set algebraically to zero by adding \texttt{mB} to the list \texttt{zero}:
\begin{verbatim}
zero=me,mmu,mU,mD,mC,mS,mB,wB,wT.
\end{verbatim}
Furthermore the number of light quarks \texttt{Nf} has to be set equal
to 5.

The tag \texttt{symmetries} specifies some further symmetries in the
calculation of the amplitudes. The information is used when the list
of helicities is generated. Possible options are:
\begin{itemize}
\item \texttt{flavour}: does not allow for flavour changing
  interactions. When this option is set, fermion lines 
  are assumed not to mix.
\item \texttt{family}: allows for flavour-changing interactions only
  within the same family. When this option is set, fermion lines 
  1-6 are assumed to mix only within families. This means that e.g.~a
  quark line connecting an up with down quark would be considered, while a
  up-bottom one would not.
\item \texttt{lepton}: means for leptons what ``flavour'' means for
  quarks.
\item \texttt{generation}: means for leptons what ``family'' means for
  quarks.
\end{itemize}
Furthermore it is possible to fix the helicity of particles. This can be
done using the command \texttt{\%<n>=<h>}, where $<n>$ stands for a PDG
number and $<h>$ for an helicity. For example \texttt{\%23=+-} specifies
the helicity of all $Z$-bosons to be ``+'' and ``-'' only (no ``0''
polarisation).

The filter
\begin{verbatim}
filter.nlo=lambda d: (not (d.isScaleless()))
\end{verbatim}
removes possible scaleless loop diagrams which may be generated by
\texttt{QGRAF}. Several predefined filters to select only subsets of
diagrams exist and can be used in this tag. The full list and some
examples can be found in~\cite{gosamhome}.



\providecommand{\href}[2]{#2}\begingroup\raggedright\endgroup

\end{document}